\documentclass[a4paper,11pt]{article}
\pdfoutput=1 
\usepackage{jcappub} 

\usepackage[normalem]{ulem}

\usepackage{mathtools}
\usepackage{cancel}
\usepackage{epsfig}
\usepackage{graphics}
\usepackage{bm}
\usepackage{color}
\usepackage{colortbl}
\usepackage{dcolumn}   
\usepackage{bm}     
\usepackage{bbm}       
\usepackage{amssymb}  
\usepackage{amsmath}
\usepackage{latexsym}
\usepackage{float}
\usepackage{ifthen}
\usepackage{caption,subfig}
\usepackage{enumerate}
\usepackage{url}
\usepackage{caption,subfig}
\usepackage{amsopn}
\usepackage{hyperref}
\usepackage{comment}
\usepackage{animate}

\usepackage{tabularx}
\usepackage{amsfonts}
\usepackage{multirow}
\usepackage{array}
\usepackage{booktabs}
\usepackage{rotating}

\usepackage{ulem}
\def\clap#1{\hbox to 0pt{\hss#1\hss}}

\def\bea{\begin{eqnarray}}
\def\eea{\end{eqnarray}}
\def\be{\begin{equation}}
\def\ee{\end{equation}}

\newcommand{\bpm}{\begin{pmatrix}}
\newcommand{\epm}{\end{pmatrix}}

\newcommand{\lp}{\left(}
\newcommand{\rp}{\right)}

\renewcommand{\geq}{\geqslant}

\newcommand{\mbA}{\mathbb{A}}
\newcommand{\tspa}{T_{j}}

\newcommand{\phwkb}{\Delta\Psi}

\title{Modified gravitational wave propagation with higher modes and its degeneracies with lensing}
\affiliation[a]{Kavli Institute for Cosmological Physics and Enrico Fermi Institute, The University of Chicago, Chicago, IL 60637, USA.}
\affiliation[b]{Department of Astronomy \& Astrophysics, The University of Chicago, Chicago, IL 60637, USA}
\affiliation[c]{Department of Physics and Astronomy, Columbia University, New York, NY 10027, USA}

\author[a,1]{Jose Maria Ezquiaga,\note{NASA Einstein Fellow}}
\emailAdd{ezquiaga@uchicago.edu}

\author[a,b]{Wayne Hu,}
 \emailAdd{whu@background.uchicago.edu}
 
\author[c]{Macarena Lagos,}
\emailAdd{m.lagos@columbia.edu}

\author[a,b]{Meng-Xiang Lin,}
\emailAdd{mxlin@uchicago.edu}

\author[a,b]{Fei Xu}
\emailAdd{feixu@uchicago.edu}

\abstract{
Low-energy alternatives to General Relativity (GR) generically modify the phase of gravitational waves (GWs) during their propagation. 
As detector sensitivities increase, it becomes key to understand how these modifications affect the GW higher modes and to disentangle possible degeneracies with astrophysical phenomena. 
We apply a general formalism---the WKB approach---for solving analytically wave propagation in the spatial domain with a modified dispersion relation (MDR). 
We compare this WKB approach to applying a stationary phase approximation (SPA) in the temporal domain with time delays associated to the group or particle velocity. 
To this end, we extend the SPA to generic signals with higher modes, keeping careful track of reference phases and arrival times. 
We find that the WKB approach coincides with the SPA using the group velocity, in agreement with the principles of wave propagation.  
We then explore the degeneracies between a GW propagation with an MDR and a strongly-lensed GW in GR, since the latter can introduce a frequency-independent phase shift which is not degenerate with source parameters in the presence of higher modes. 
We find that for a particular MDR there is an exact degeneracy for wave propagation, unlike with the SPA for particle propagation.  
For the other cases, we search for the values of the MDR parameters that minimize the $\chi^2$ and conclude that strongly-lensed GR GWs could be misinterpreted as GWs in modified gravity. 
Future MDR constraints with higher mode GWs should include the possibility of frequency-independent phase shifts, allowing for the identification of modified gravity and strong lensing distortions at the same time. 
}
\date{\today}

\begin{document}

\maketitle

\section{Introduction}
The LIGO--Virgo--KAGRA Collaboration (LVK) \cite{Aasi2015,Acernese_2014,KAGRA:2020tym} has already confirmed the detection of 90 gravitational wave (GW) events \cite{LIGOScientific:2021djp} from binary compact coalescences. Most of these events correspond to nearly equal-mass binary black holes (BBHs) with no precession, and thus exhibit simple waveforms that are well described by quadrupole radiation with dominant spherical harmonic numbers $(\ell=2, m=\pm 2)$. However, waveforms from unequal-mass or precessing binaries will have measurable energy emission in higher harmonics, as it has already been the case for a few of the GW events, see e.g.\ \cite{LIGOScientific:2020stg,Abbott:2020khf} for LVK detections and \cite{Olsen:2022pin,Nitz:2021zwj} for independent groups catalogs. Although these events are currently rare, they will be detected more often, especially in the next-generation of GW detectors \cite{LISA:2017pwj, Maggiore:2019uih, Evans:2021gyd} when the typical signal-to-noise ratio of the events will be dramatically increased.

In general, the higher harmonics contain crucial information to test the behavior of gravity in both the strong and weak-field regime. 
In particular, if the GW emission is according to general relativity (GR) and the GWs propagate in a perfectly homogeneous and isotropic universe, the frequency and phases of each harmonic of the waveform will have a well defined evolution dictated by the dynamics of the binary. 
However, modifications during emission and/or propagation could introduce distortions in the observed signal, leaving thus a detectable non-trivial signature.
For instance, if GWs suffered from gravitational lensing during their propagation, it has been shown that the waveform will suffer a constant phase shift, which will change the fiducial phase relation between harmonics expected in unlensed GR signals \cite{Dai:2017huk,Ezquiaga:2020gdt}. As a result, a GW signal containing higher harmonics will arrive distorted at the detector. 
In addition, modifications of the emission process will also affect the fiducial frequency evolution expected in GR and lead to distorted detected signals \cite{Yunes:2009ke, Tahura:2018zuq}. Depending on the details of the type of distortions and the detector sensitivity, it may occur that these distortions become indistinguishable from other astrophysical effects, which poses a challenge to disentangle various physical effects.  

In this paper we consider the modified propagation of GWs caused by a modified dispersion relation (MDR), assuming that the emission process is unchanged with respect to GR---motivated by low-energy modified gravity theories. 
We discuss how an MDR can affect a GW signal with higher modes, and how it can mimic similar signatures to those of strong gravitational lensing. Specifically, we first apply and compare two different analytical approaches to take into account a modified dispersion relation in GW signals: the WKB formalism \cite{Jimenez:2019lrk} based on spatial propagation of wavenumbers in the wavepacket, and the stationary-phase approximation (SPA) formalism based on temporal propagation of frequencies in the wavepacket \cite{Will:1997bb,Mirshekari:2011yq,LIGOScientific:2019fpa}. In \cite{Ezquiaga:2021ler}, some of the authors of this paper used the WKB approach to analyze the effect of various MDRs on GWs with a dominant quadrupole radiation, and in this paper we apply the same WKB approach to GWs with higher multipoles. We highlight that this WKB approach is general and systematic, and can be used to obtain analytic expressions for the distortions to any emitted signal in frequency-domain, irrespective of its complexity (e.g.\ containing higher multipoles, precession, eccentricity, etc). We also compare our WKB approach to the more common approach previously used in the literature, the SPA, which has been mostly applied to GWs with a single multipole mode. Indeed, previous LVK data analyses \cite{LIGOScientific:2019fpa,Abbott:2020jks,LIGOScientific:2021sio} have left aside all events with evidence of higher multipoles or have analyzed them under the assumption of quadrupolar radiation. In this paper, we thus generalize the SPA formalism to GWs with higher multipoles. 
Also, the SPA approach can be used assuming that GWs propagate at either the group or particle velocity, and here we show that the WKB approach is equivalent to the SPA only in the former case. We illustrate the differences and similarities of both approaches with explicit examples of GW signals from unequal mass BBHs. The results presented here show how to properly account for a modified propagation of GWs when analyzing future detections with higher modes.

As a practical use of the WKB approach, we also analyze strong gravitational lensing of GWs in GR, and show how future GW analyses of waveforms with higher multipoles could wrongly show evidence for a modified propagation if lensing is not taken into account in the waveform templates.
More precisely, we compare the $\chi^2$ goodness-of-fit measure of unlensed GR templates versus MDR templates to a strongly-lensed GR GW signal from unequal mass binaries.
We find that for inclined binaries with very low mass ratios $q=m_2/m_1\lesssim 0.1$ or events with very high signal-to-noise, there can be various MDRs that provide a better fit to the signal for an A+ LIGO detector \cite{Aasi:2013wya}---including a special case where the degeneracy between the MDR and strong lensing is exact---and hence in any of these events strong-lensing distortions could be misidentified as evidence for modified gravity. 
However, as $q$ approaches unity or the inclination approaches zero (or $\pi$), the strong-lensing effects can be well mimicked by changes in source parameters of unlensed GR templates, and will not be mistaken for an MDR.  
With future GW detectors such as Einstein Telescope \citep{2020JCAP...03..050M} and Cosmic Explorer \citep{2019BAAS...51g..35R},  
the expected number of detected lensed events may increase by $\sim$ 2 orders of magnitude compared to advanced LIGO \citep{2018MNRAS.476.2220L, 2019arXiv190403187T, 2021arXiv210514390X} while not being explicitly pre-identified as lensing events through the association of multiple images \cite{Caliskan:2022wbh}.  
The number of events with measurable higher multipoles also increases, thanks to the larger signal-to-noise ratios (SNR). 
The best way to avoid misinterpreting lensed signals in the future is to use strongly lensed templates in parameter estimations (and ideally in searches as well).
Therefore, our results highlight the importance of lensed templates, especially for high SNR events whose fits are sensitive to small differences between templates.

This paper is structured as follows. In Section \ref{sec:WKB} we summarize the main features of the WKB approach applied to the modified propagation of GWs. In Sec.\ \ref{sec:spa} we extend the SPA approach to GWs with higher multipoles, discuss the difference between using group and particle velocity, and finally compare to the WKB approach. In Sec.\ \ref{sec:lensing} we focus on lensed GWs, and apply the WKB approach to make templates of GWs with MDRs that are compared to unlensed GR templates. We quantify the $\chi^2$ goodness of fit of these templates for various mass ratios. Finally, in Sec.\ \ref{sec:conclusions} we summarize our results and discuss their consequences.
Throughout the paper we work in units of $\hbar=1$, and define $c$ as the speed of light.

\section{WKB approach}\label{sec:WKB}

In \cite{Ezquiaga:2021ler} some of the authors of this paper developed an analytical approach for describing the possible modified propagation of GWs on a cosmological background, and applied it to the case of GWs interacting with an additional tensor field, extending the work of \cite{Jim_nez_2020}. However, the approach is general and also applicable to models in which GWs do not interact with extra tensor fields, but still have a modified propagation (either because the self-interactions of gravity are different to those in GR, or because GWs interact with cosmological fields that do not carry tensor degrees of freedom). The approach relies on the WKB approximation (for cases in which the GW period is much smaller than Hubble time), and it mainly consists of three steps:
\begin{enumerate}
    \item  For any given model, write down the equation of motion for the GW field $h$ in $k$-space, where the coefficients in the equation are allowed to evolve on cosmological timescales. In this case, $h$ is a function of conformal time $\eta$ and comoving wavenumber $k$.
    \item Solve the equation using the WKB approach to obtain the propagation of GWs from emission $h(k,\eta_e)$ to detection $h(k,\eta_o)$ of each wavenumber. The solution was found to look like a plane wave with a time-dependent phase determined by the dispersion relation $\omega(k)$, with $\omega$ being the conformal frequency of the wave. If the equations of motion lead to a dispersion relation different to GR, i.e.\ $\omega\not= ck$, then GWs will suffer cumulative phase distortions during their propagation, compared to GR, and these distortions have a simple analytical expression.
    \item Motivated by current GW observations, we assume that the deviations from GR in the dispersion relation are small. In that case, we can locally translate the signal detected in $k$-space $h(k,\eta_o)$ to that in frequency space $h(\omega,x_o)$, with this latter determining the temporal profile of GWs detected at a given location $x_o$ (as in observations).
\end{enumerate}

Note that in Step 1 one can make arbitrary choices of the coefficients in the equation of motion for $h$, which will always be consistent with some quadratic cosmological lagrangian for $h$.  Whether this lagrangian will have a consistent and stable non-linear gravitational action associated is not known, and we will not restrict ourselves to equations from currently known non-linear modified gravity models in the literature. Instead, we take a more phenomenological approach and consider arbitrary choices in the equation of motion for $h$. Furthermore, instead of considering a given equation of motion (as mentioned in Step 1), we will use the result obtained in Step 2 and consider a family of parametrized modified dispersion relations (MDRs) that distort the phase evolution of GWs compared to GR. The effect of parametrized MDRs has been previously studied in the literature and its observational effects on the phase evolution of GWs have been calculated using the stationary-phase approximation (SPA) approach mostly for quadrupolar radiation during the inspiral, see e.g.\ \cite{Will:1997bb,Mirshekari:2011yq}, but also some eccentric cases \cite{Yunes:2009yz}. 
In this paper, we instead use the WKB approach we developed in \cite{Ezquiaga:2021ler}, and discuss its comparison to the SPA approach in the next section, extending also the SPA formalism to signals with arbitrary multiple frequency modes. 

Let us begin by discussing explicitly how we use the WKB approach to obtain the observed GWs with an MDR. 
As mentioned, we will ignore Step 1 and only use the results of Step 2 and Step 3. These last two steps are actually independent from each other and we can discuss them separately. Regarding Step 3, suppose that we are given a wave in $k$-space $h(k)$. The spherical wave propagating along a direction $x$ can be locally represented as plane waves around the detector at $x_o=0$ and reception time $T=t-t_c$ where $t_c$ is some arbitrary reference time, which we fix as the arrival of the coalescence frequency in GR. In this case, any of the polarizations of the wave can be expressed as\footnote{Note that here $h$ does not represent the total GW strain defined typically in the literature, which includes observational effects of the detector (e.g.\ the antenna pattern function). Instead, $h$ is the generic name that we give to any of the two polarizations of the wave that only includes source and propagation effects, but not detector effects.}
\begin{equation}
h(x,T) = 2\int_0^\infty \frac{dk}{\sqrt{2\pi}}\Re\left[ e^{i(k x-\omega T)} h(k) \right],
\end{equation}
where we assume $k,\omega>0$ and the wave propagates in the $+x$ direction. In addition, if the MDR deviations from GR are small, near detection in both space and time, we can approximate 
$\omega(k) \approx ck$ and consider the phase distortions due to the MDR to accumulate only over cosmological distances.
Therefore, in order to get the temporal waveform locally at the detector we can proceed as follows:
\begin{equation}\label{eq:k_to_w}
h(0,T) =  2\int_0^\infty \frac{dk}{\sqrt{2\pi}} \Re\left[ e^{-i \omega T} h(k)\right]  \approx   2 \int_0^\infty \frac{d\omega}{\sqrt{2\pi}}  \Re\left[ e^{-i \omega T} h(\omega) \right].
\end{equation}
From here we can read off that the frequency-domain waveform, defined relative to $T=0$, is given by $h(\omega)=ch(k)$.
This equivalence between the frequency and wavenumber domains applies exactly for GR---or what we will define as the fiducial waveform $h_{\rm fid}$---and to good approximation {\it locally} for an MDR. The effect of the MDR is that $h(k)$ can change in a manner different from GR  across cosmological time scales and, therefore, in order to apply Eq.\ (\ref{eq:k_to_w}) we will need to input the waveform of $h(k)$ at the detection time. 
This last part is done in Step 2, where one calculates the change in $h(k)$ from emission to detection. In \cite{Ezquiaga:2021ler}, under the WKB approximation, it was obtained that, for a source at redshift $z_s$ with $1+z_s=1/a(t_e)$ where $a(t_e)$ is the scale factor at the emission time, the $k$-space waveform suffers a phase correction due to $\Delta \omega = \omega(k)-ck\not=0$. This phase correction is such that:
\begin{equation}
\label{hsol_k}
h(k) = h_{\rm fid}(k) e^{-i \int_0^{z_s} dz\, \Delta\omega /H(z)} ,
\end{equation}
where $h_{\rm fid}(k)$ is the waveform that would be detected in GR, and the phase correction is described by an integral of $\Delta \omega$,  which takes into account the possibility that the coefficients in the dispersion relation may evolve on cosmological times.
From Eq.\ (\ref{hsol_k}) we can 
read off that the frequency-domain waveform at the detector also acquires the same phase shift 
\begin{equation}
h(\omega) = h_{\rm fid}(\omega)  e^{-i \int_0^{z_s} dz\, \Delta\omega /H(z)}  \equiv h_{\rm fid}(\omega)   e^{i \Delta\Psi}. \label{hsol_omega}
\end{equation}
Note that in Eq.\ (\ref{hsol_k}) we have assumed that the only modification in the propagation of the wave is a phase distortion due to an MDR. 
In more general scenarios, there could also be changes in the amplitude---for example due to an extra friction term in the propagation equation---as well as changes of polarization---due to chiral interactions that break left-handed and right-handed polarization symmetry. Both of these effects were considered in the formalism in \cite{Ezquiaga:2021ler}, and could be easily taken into account in this approach, but for simplicity they will not be considered in this paper since we are mostly interested on distortions of the phase. 
Also, we note from Eq.\ (\ref{eq:k_to_w}) that a frequency-domain phase shift $h(\omega) \rightarrow h(\omega) e^{i\Delta\Psi}$ is equivalent to a time-domain shift of the opposite sign $ T \rightarrow  T -\Delta\Psi/\omega$, which will be useful to recall below.

For concreteness, in this paper we consider an MDR that is parametrized by a single power-law correction
\begin{equation}\label{eq:MDR}
\omega^2 = c^2k^2 \left[1 + 
\mbA \left( \frac{ck}{a} \right)^{\alpha-2}\right]\,,
\end{equation}
where $\mbA$ is assumed to be constant and $k/a$ converts the comoving wavenumber (non-redshifting) to physical (redshifting) wavenumber. At leading order in $\mbA$ we have
\begin{equation}
\frac{\Delta \omega}{\omega}  \approx \frac{\mbA }{2} \left( \frac{ck}{a}\right)^{\alpha-2}
\end{equation}
and therefore, the GW suffers a phase correction given by
\begin{equation}
\Delta\Psi \approx  -\int_0^{z_s} \frac{dz}{H(z)}  \frac{\mbA \omega}{2} \left( \frac{ck}{a} \right)^{\alpha-2} \approx -\frac{\mbA \omega^{\alpha-1}}{2}
 \int_0^{z_s} \frac{dz}{H(z)} (1+z)^{\alpha-2}\,.
\end{equation}
It is conventional to define  an effective distance \cite{Will:1997bb, Mirshekari:2011yq}
\begin{equation}\label{Dalpha}
D_\alpha(z_s) =  (1+z_s)^{1-\alpha}\int_0^{z_s} \frac{c dz}{H(z)} (1+z)^{\alpha-2}
\end{equation}
so that we finally have 
\begin{equation}\label{Psi_wkb}
\Delta\Psi = -\frac{\mbA [\omega(1+z_s)]^{\alpha-1}}{2c}D_\alpha
\end{equation}
where $\omega(1+z_s)$ is also the (angular) frequency at emission in physical units.
Note that $\mbA$ and $D_\alpha$ are dimensionful quantities.   
The dimensionless change in the dispersion relation scales as
${\mbA} \omega^{\alpha-2}$
and it is assumed to be small ${\cal O}(H_0/{\omega})$.

The approach described here has various notable features. On the one hand, the modification of the phase $\phwkb$ is only due to the propagation, and hence it is independent of the emission mechanism of the wave. This means that (\ref{hsol_omega}) is applicable to \emph{any} GW, including those with complicated structures such as from unequal mass  or precessing binaries of compact objects. On the other hand, 
$\phwkb$ gives the MDR phase shift of a given frequency component at a fixed time, here the arrival time of the coalescence frequency under GR, relative to the same signal propagated in GR, which is not a direct GW observable.   
For example, if $\alpha=2$, then GWs travel at a common speed that is different to $c$ and while $\Delta \Psi$ is non-vanishing, there are no distortions of the observed wave vs GR.

A more direct observable is the arrival phase and relative time delays of each frequency component for the observed MDR signal.
We shall see next that 
 for a binary inspiral waveform, relating this arrival phase to $\Delta \Psi$ involves the stationary phase approximation (SPA). However in contrast with similar SPA treatments in the literature \cite{Will:1997bb,Mirshekari:2011yq} we show that the arrival time is characterized by the group velocity $v_g = \partial \omega/\partial k$ rather than the particle velocity $v_p = c^2 k/\omega$ (see \S \ref{sec:particle_velocity}).

\section{Stationary Phase Approximation Approach}\label{sec:spa}
Previous works, e.g.\ \cite{Will:1997bb,Mirshekari:2011yq}, have employed the stationary phase approximation (SPA) in order to obtain the phase distortions of a waveform due to a modified propagation relation.\footnote{Note that the SPA is also ubiquitous in the GW literature in order to obtain analytical approximations of frequency-domain waveform with the goal of assessing their detectability. See \cite{Cutler:1994ys,Droz:1999qx} for seminal works.} In particular, if a given GW signal can be decomposed into a linear superposition of modes, each with a monotonic time evolution in detected frequency $f= \omega/2\pi>0$, then the arrival phase of a given frequency for each mode can be calculated under the SPA. For a signal dominated by a single mode, this is advantageous since the arrival time and phase is a direct observable whereas the frequency-domain phase at a common time is not.  
On the other hand, due to its construction via a single arrival time for a given frequency, the SPA would seem to require the single mode condition.  In this section, we summarize the SPA method extending it to the case of waveforms with multiple modes, and compare it to the WKB approach.

We start by considering that the 
\emph{detected} GW wave for any polarization, $h(T)$, of a compact binary coalescence is the sum of contributions from various emission modes labelled by $j$ as
\begin{align} 
    h(T) &= \sum_j \Re\left[ A_j(T)e^{-i\Phi_j(T)}\right]\,, \\
    \Phi_j(T) 
   & =
    2\pi\int_{T_{j,c}}^{T}F_j(T')dT' +\Phi_{j,c}\,, \label{eq:Phi_j}
\end{align}
where $T_{j,c}$ is the arrival time of the coalescence frequencies, $F_{j,c}$, of each mode. These coalescence frequencies are defined as the frequencies of each mode at the common emission time of the peak signal, where the emitted signal itself is assumed to be given by GR in this paper. 
Given an MDR with a frequency dependent propagation velocity, these coalescence frequencies that were emitted at the same time will not necessarily arrive at the same time, and these different arrival times are described by  $T_{j,c}$.  This can lead to a temporal decoherence of the peak signal and more generally a modified interference between modes. 
In addition, here we have
characterized the modes by their real amplitude $A_j(T)$ and phase $\Phi_j(T)$, which itself is typically parametrized relative to its value $\Phi_{j,c}$ at $T_{j,c}$, as measured by the detector. Here, the time derivative of the phases defines the frequencies $F_j$ of each mode, which depend on time. We will assume a monotonic relationship such that there is a unique inverse transformation (or at least monotonic by pieces). 

The angular distribution of the emitted GW signal is usually decomposed into spin-2 spherical harmonics, labelled by the multipole of degree $\ell$ and order $m$. In the case of binary systems without precession (due to spins or eccentricity), each spherical harmonic will contribute with a GW mode that has a different frequency evolution $F_j(T)$, and thus each mode $j$ will label each pair of harmonic numbers $(\ell,|m|)$. For precessing systems, each spherical harmonic can contribute to the GW source with various components at different frequencies, and in these cases additional labels have to be introduced to distinguish the different $j$ modes. 
As a concrete example of the frequency evolution, in the simplest case of a GW signal from an equal-mass binary in a nearly circular non-precessing orbit, the waveform is dominated by the mass quadrupolar radiation $(\ell=2, |m|=2)$, or $j=22$, and there is only one frequency mode $2 \pi F_{22}=2\omega_\text{orb}$ at twice the orbital frequency, whose time evolution during the inspiral in GR is given by \cite{Maggiore:1900zz}
\begin{equation} \label{eq:chirp_f}
    \frac{dF_{22}}{dT}=\frac{96}{5}\pi^{8/3}\mathcal{M}_z^{5/3}F_{22}^{11/3}\,, 
\end{equation}
where 
\begin{equation}
\mathcal{M}_z\equiv (1+z)\frac{(m_1m_2)^{3/5}}{(m_1+m_2)^{1/5}}
\label{eq:chirpmass}
\end{equation}
 is the redshifted chirp mass of a binary system with masses $m_1$ and $m_2$.   
In this section, we will not assume any specific GW source, and keep the discussion general for each mode $j$.

We continue now by following the procedures described in \cite{Will:1997bb} to work out the phase shift that GWs suffer in the frequency domain due to an MDR  during propagation. Going to frequency domain,
\begin{equation}\label{fourier_f_to_t}
  h(f) =   \int_{-\infty}^\infty   \frac{dT}{\sqrt{2\pi}}  e^{i 2\pi f T} h(T) .
\end{equation}
We can solve the Fourier transform analytically using the SPA for each mode (valid when their amplitudes evolve in time much slowly than their phases). 
Under the SPA, one obtains that  $h(f)$ is given by
\begin{align}
    h(f)&=\sum_j\frac{A_j(\tspa)}{2\sqrt{2\pi}\sqrt{dF_j/dT(\tspa)}}e^{i\Psi_j(f)}\,, \label{h_spa}\\
    \Psi_j(f) &= 2\pi f\tspa(f)-\Phi_j(\tspa(f))-\pi/4\,, \label{eq:Psi_j}
\end{align}
where $\tspa(f)$ is the arrival time of the stationary point, which is found solving
\begin{equation} \label{eq:freq_from_phase}
    \left.\frac{d\Phi_j(T)}{dT}\right\vert_{T=\tspa}=2\pi F_j(\tspa)\equiv 2\pi f\,.
\end{equation}
Note that due to the definition in Eq.\ (\ref{fourier_f_to_t}), $\Psi_j(f)$ describes the frequency-domain phase of any frequency component at $T=0$ as in the previous section, not at its arrival time $T_j(f)$. 
Therefore, we can  interpret Eq.\ (\ref{eq:Psi_j}) as the sum of two components:
$-\Phi_j(\tspa(f))-\pi/4$, the frequency-domain phase at arrival at the detector
and $ 2\pi f\tspa(f)$, the phase difference corresponding to shifting from the arrival time $\tspa(f)$ to a common time, given by $T=0$.
We will hereafter use the shorthand notation $\Phi_j(f) \equiv \Phi_j(T_j(f))$ and consider it as a contribution to the frequency-domain phase.

Explicitly, this term  can be computed as:
\begin{equation}
\begin{split}
    \Phi_j(f)&=2\pi\int_{T_{j,c}}^{\tspa(f)}F_j(T')dT' +\Phi_{j,c} \\
    &=2\pi\left(F_j(\tspa)\tspa-F_j(T_{j,c})T_{j,c} - \int_{F_{j,c}}^{F(\tspa)}T_j(F')dF'\right) +\Phi_{j,c} \\
    &=2\pi\left(f\tspa-F_j(T_{j,c})T_{j,c} - \int_{f_{j,c}}^{f}T_j(f')df'\right) +\Phi_{j,c} \\
    &=2\pi\left(f\tspa-fT_{j,c} - \int_{f_{j,c}}^{f}(T_j(f')-T_{j,c})df'\right) +\Phi_{j,c} \,,
\end{split}
\label{eq:Phij}
\end{equation}
where $T_{j,c}=T_j(f_{j,c})$ and in the second line we have integrated by parts assuming that for each mode the relation $F_j(T)$ can be inverted (at least by pieces). In the third line we have used the definition of the stationary point and we have changed the integrand to make explicit the dependence in $f$. 
Now, applying this result to the frequency-domain phase we obtain
\begin{equation}\label{eq:fourier_phase}
    \Psi_j(f) = 2\pi fT_{j,c}+2\pi\int_{f_{j,c}}^{f}(T_j(f')-T_{j,c})df' - \Phi_{j,c} - \pi/4.
\end{equation}
So far, this is the standard GR approach (see e.g.~\cite{Cutler:1994ys}), where the integrand in (\ref{eq:fourier_phase}) is the time delay of different frequency modes with respect to the coalescence time.
We emphasize that, in this approach, all the quantities are given with respect to detection: $f$ is the detected frequency, $\Phi_{j,c}$ is the detected  phase of coalescence of a mode $j$, and $T_{j,c}$ is the detected arrival time of the coalescence frequencies. 
If there is a modified propagation of GWs, $\Phi_{j,c}$ can be different from the coalescence phase at emission whenever the phase velocity is different from the physical propagating velocity associated to the coalescence frequency. In addition, we note that, since an MDR during propagation changes the arrival time of different frequencies, the frequency-time evolution will be modified in the detector frame $(f,T)$ with respect to the relation at emission $(f_s,\Delta t_e)$. In the next subsections, we take these two effects into account.

In calculating the travel time of different frequency components, it is crucial to establish at which speed the GW signal is propagating.   Given that the arrival phase is calculated under the SPA, the  speed to associate with the arrival of a given frequency in the wavepacket is the group velocity defined by the MDR. This will be the assumption made in Subsec.\ \ref{sec:group_velocity} and shown to be consistent with the WKB prediction as expected. However, in Subsec.\ \ref{sec:particle_velocity}, we will also consider a common approach in the literature which associates the particle velocity to the arrival time. 

%------
%Group velocity
%------
\subsection{Group velocity} \label{sec:group_velocity}

To evaluate 
the frequency domain phase in Eq.~(\ref{eq:fourier_phase}), 
we need to know the arrival time in the SPA for a given frequency component in the signal. This is determined by the group velocity $v_g=\partial\omega/\partial k$,
which at linear order in $\mbA$ simplifies to 
\begin{equation}\label{eq:vg_exp}
    \frac{v_g}{c}
    \approx 1 - \frac{1}{2}(1-\alpha)\mbA\lp 2\pi f\rp^{\alpha-2}\,,
\end{equation}
where $2\pi f=\omega/a$ is the physical frequency.  
For example, a common model considered in the literature is massive gravity, where $\alpha=0$ and $\mbA=m^2c^4>0$ with $m$ being the mass of the graviton. In this model, the group velocity becomes: 
\begin{equation}
    \frac{v_g}{c}\approx 1 - \frac{1}{2}\frac{m^2c^4}{\lp 2\pi f\rp^{2}}\,,
\end{equation}
where one can see that waves always propagate slower than the speed of light, and that the MDR is dispersive, with lower frequency modes propagating slower than high frequency modes. 
Another special case is when $\alpha=1$, since the first-order correction in Eq.\ (\ref{eq:vg_exp}) vanishes: GWs propagate at the speed of light for $\alpha=1$. 
Finally for $\alpha=2$, the group velocity is constant but differs from that of light.

Next, in order to obtain the frequency-domain phase of GWs (see  Eq.\ (\ref{eq:fourier_phase})) for a given dispersion relation, we need to calculate the time delay of any frequency mode with respect to the coalescence frequency. From Eq.\ (\ref{eq:vg_exp}), we find that the detected time delay between two different frequencies $\Delta t =\Delta T= T(f)-T(f')$, at leading order in $\mbA$,  is 
\begin{equation} \label{eq:dt_vg}
    \Delta t = (1+z_s)\left(\Delta t_e +\frac{(1-\alpha)}{2}\mbA D_\alpha\left(\frac{1}{(2\pi f_s)^{2-\alpha}}-\frac{1}{(2\pi f_s^{\prime})^{2-\alpha}}\right)\right)\,,
\end{equation}
where $D_\alpha$ is given by Eq.\ (\ref{Dalpha}), $\Delta t_e$ is the intrinsic emission time delay, and we have referenced the observed frequencies to the corresponding source-frame frequencies $f_s$ and $f_s'$ assuming that the source is at redshift $z_s$ and the observer is at $z=0$. 
As a pedagogical note, we remind the reader that Eq.\ (\ref{eq:dt_vg}) can be computed equating the comoving distance between two signals emitted at time $t_e$ and $t_e-\Delta t_e$ with comoving angular frequencies $\omega$ and $\omega'$ respectively. More explicitly, this means
\begin{equation}\label{same_chi}
    \chi=\int_{t_e}^{t}v_g(\omega,t)\frac{dt}{a}=\int_{t_e-\Delta t_e}^{t-\Delta t}v_g(\omega',t)\frac{dt}{a}\,,
\end{equation}
where $t$ and $t- \Delta t$ are the arrival time of the signals. For small time delays, compared to Hubble time, Eq.\ (\ref{same_chi}) can be approximated to 
\begin{equation}\label{eq:dt_vg_detail}
\begin{split}
   \Delta t &=(1+z_s)
   \frac{v_{g}(\omega',t_e)}{v_g(\omega',t)}\Delta t_e - \frac{a(t)}{v_g(\omega',t)}\int_{t_e}^{t}\lp v_g(\omega,t')-v_g(\omega',t')\rp\frac{dt'}{a} \\
   &\simeq (1+z_s)\lp \Delta t_e - \frac{1}{(1+z_s)}\int_{t_e}^{t}\frac{v_g(\omega,t')-v_g(\omega',t')}{c a(t')}dt' \rp ,
\end{split}
\end{equation}
when assuming that $v_g\approx c$. Note that we have dropped corrections of order $\Delta t_e \mbA$ since MDR effects only accumulate significantly across the cosmological propagation distance. When using Eq.\ (\ref{eq:vg_exp}), and re-expressing this last integral in terms of the source frequencies $2\pi f_s=\omega/a(t_e)$ and $2\pi f_s'=\omega'/a(t_e)$, one obtains Eq.\ (\ref{eq:dt_vg}).

Now that we have the time delay (\ref{eq:dt_vg}), we incorporate it into the phase using Eq.\ (\ref{eq:fourier_phase}). We find, at leading order in $\mbA$, the frequency-domain phase of the signal (for $f_s>0$):
\begin{align}\label{eq:fourier_phase_vg}
        \Psi_{j}(f) 
        ={}& 2\pi f T_{j,c} + 2\pi\int_{f_{sj,c}}^{f_s}\!\!\! \lp t_{e,j}(f'_s)-t_{e,j}(f_{sj,c}) \rp df_s' \left. -\frac{\mbA}{2} D_\alpha \! \left[  \frac{1}{(2\pi f_s')^{1-\alpha}}+\frac{ (1-\alpha)2\pi f'_s}{(2\pi f_{sj,c})^{2-\alpha}} \right] \right\rvert_{f_{sj,c}}^{f_s} \nonumber\\ &
        - \Phi_{j,c} - \frac{\pi}{4}, 
\end{align}
where the integral term depends on the emission times $t_{e,j}$ of each mode for a source-frame frequency $f_s$ and the coalescence frequencies $f_{sj,c}$. This result applied to $\alpha=0$ for a single mode agrees with the result of (3.6) in \cite{Will:1997bb}. In Eq.\ (\ref{eq:fourier_phase_vg}) we can see that the MDR dependent contributions can be grouped into the last 3 terms of 
\begin{align}
     \Psi_{j}(f)  ={}&  
     2\pi\int_{f_{sj,c}}^{f_s}\lp t_{e,j}(f'_s)-t_{e,j}(f_{sj,c}) \rp df_s' - \frac{\pi}{4}  + 2\pi f \left[ T_{j,c}  - (1-\alpha)\frac{\mbA}{2}  \frac{(1+z_s)D_\alpha}{(2\pi f_{sj,c})^{2-\alpha}} \right] \nonumber \\
     {}&  - \left[ \Phi_{j,c} -  (2-\alpha)\frac{\mbA}{2} \frac{D_\alpha}{(2\pi f_{sj,c})^{1-\alpha}}\right]  + \Delta \Psi(f) \,,\label{eq:psi_tot}
\end{align}
where
\begin{equation}
    \Delta \Psi (f)=- \frac{ \mbA }{2} \frac{D_\alpha}{(2\pi f_s)^{1-\alpha}}\label{Psi_spa}
\end{equation}
which is the same for all components $j$ of the GW signal.

We have grouped the terms in this way to highlight that 
$\Delta\Psi$ is the same term that was found as the phase correction with the WKB formalism in Eq.\ (\ref{Psi_wkb}). As we will see next, all the other terms that depend on $\mbA$ in Eq.\ (\ref{eq:psi_tot}) cancel out in such a way that $\Delta \Psi$ encodes all the phase corrections to GR due to an MDR in the frequency-domain phase $\Psi_j$ at $T=0$. 

In the first square bracket of Eq.\ (\ref{eq:psi_tot}) we have terms proportional to the frequency that simply describe time delays of the GW signal. In particular, $T_{j,c}$ is the time difference in the arrival times of $f_{j,c}$ in the MDR model compared to GR. This can be calculated explicitly following an analogous derivation to Eq.\ (\ref{same_chi})-(\ref{eq:dt_vg_detail}) but now for the same frequency $f_{j,c}$ that is emitted at the same time, but there are two different velocities to compare: $c$ and $v_g(\mbA)$. If we do this, we obtain that:
\begin{equation}
\label{eq:Tjc}
    T_{j,c} = (1-\alpha)\frac{\mbA}{2}  \frac{(1+z_s)D_\alpha}{(2\pi f_{sj,c})^{2-\alpha}}
\end{equation}
so that the first square bracket of Eq.\ (\ref{eq:psi_tot}) vanishes.    
We note that only relative shifts between the coalescence time of different modes can potentially be observed in the GW waveform through their potential distortions due to interference effects. 
However, if the GW source also emits an EM signal that is detected, comparison in their arrival times may be used to constrain the absolute GW time delay due to its modified propagation. GW170817~\cite{LIGOScientific:2017vwq} can be used for this test, although this source had very low redshift, which makes the derived constraints weaker than the ones obtained from the phase evolution analysis of the population of BBHs \cite{Abbott:2020jks}. In particular, in the case of a massive graviton ($\alpha=0$), the GW upper bound from BBHs population was found to be $m < 1.76\times 10^{-23}\text{eV}/c^2$ (or $\mbA<1.37\times 10^{-45}\text{eV}^2$) at $90\%$ confidence level, whereas from GW170817 the constraint was $m<9.51\times 10^{-22}\text{eV}/c^2$.

Regarding the second square bracket of Eq.\ (\ref{eq:psi_tot}),
we recall that the SPA approach defines all phases at detection. However, the detected coalescence phase $\Phi_{j,c}$ could be different from the emitted one whenever the phase velocity does not equal the propagation speed, in this case the group velocity. 
In particular, we can relate the emitted and observed phase by
\begin{equation}
    \Phi_{j}(f) = \Phi^{e}_{j}(f) + \Delta\Phi (f) \,,
    \label{eq:Phijc_group}
\end{equation}
where $\Phi^{e}_{j}$ is the emitted phase and $\Delta\Phi$ is given by
\begin{equation}\label{eq:DeltaPhi}
    \Delta\Phi (f) = \int_{t_e}^{t} \left[ v_\text{ph}(t') - v_g(t')\right]  \frac{k}{a(t')}dt'
\end{equation}
for a general frequency $f$ emitted at $t_e$ and detected at $t$.
For the dispersion relation under consideration, we can calculate $\Delta\Phi$ explicitly. The phase velocity at leading in $\mbA$ is given by 
\begin{equation}\label{vphase_linearA}
    \frac{v_\text{ph}}{c}=\frac{\omega}{c k}=\sqrt{1+\mbA (c k/a)^{\alpha-2}}\approx  1 + \frac{1}{2}\mbA (c k/a)^{\alpha-2}\,,
\end{equation}
which can be compared to the group velocity in Eq.\ (\ref{eq:vg_exp}). 
At the arrival time, the change in the phase compared to the initial phase,  
for a given frequency $f$, is then given by:
\begin{eqnarray}
\Delta\Phi(f) 
&\approx & (2-\alpha) \frac{ \mbA }{2}\frac{D_\alpha}{(2\pi f_{s})^{1-\alpha}} ,
\label{eq:DeltaPhi_vg}
\end{eqnarray}
where we have used that $f=f_s/(1+z)$. As we can see from here,  this term is the same as that one in Eq.\ (\ref{eq:psi_tot}) at the coalescence frequency. Finally, Eq.\ (\ref{eq:psi_tot}) then simplifies to
\begin{align}
        \Psi_{j}(f) & =  \Psi_j^{\rm GR}(f) + \Delta \Psi(f)   \, ,\label{eq:psi_tot2}
\end{align}
at leading-order in $\mbA$, 
where 
\begin{equation}\label{eq:Psi_GR}
    \Psi_j^{\rm GR}(f) = 2\pi\int_{f_{sj,c}}^{f_s}\lp t_{e,j}(f'_s)-t_{e,j}(f_{sj,c}) \rp df_s'- \Phi^e_{j,c} - \frac{\pi}{4}.
\end{equation}

From here, we indeed confirm that $ \Delta \Psi(f)$ is the only modification to GR that appears due to the MDR, which agrees with the WKB approach described in the previous section.
Finally notice that evaluating the SPA expression for $\Phi_j(f)$ in Eq.~(\ref{eq:Phij}) with the time delay (\ref{eq:dt_vg}) and observed SPA phase at the coalescence frequency (\ref{eq:Phijc_group}) we obtain that the observed SPA phase 
\begin{equation}
\Phi_j(f) = \Phi_j^{e}(f) + \Delta\Phi(f)
\end{equation}
for all frequencies, which checks the self consistency of the phase propagation formula in Eq.~(\ref{eq:DeltaPhi}).
This frequency-domain phase shift from emission $\Delta \Phi(f)$, and the frequency-dependent arrival time $\Delta t(f,f_{j,c})$ relative to $f_{j,c}$ are more direct observables than $\Delta\Psi(f)$, the total phase change relative to GR at the fixed time of the latter.
For example, for $\alpha=2$, GWs simply propagate at a different speed than light and the MDR contribution to $\Delta t$ is constant in frequency and $\Delta \Phi(f)=0$.   Unlike $\Delta\Psi$, the quantities $\Delta t$ and $\Delta \Phi$ directly reflect the fact that there are no distortions of the waveform for this case.  On the other hand, for any given $\alpha$ the two are simply related as
\begin{equation}
\frac{\Delta\Phi}{\Delta\Psi} = \alpha-2\,,
\label{eq:phaseratio}
\end{equation}
independently of frequency.

Conversely note that when $\alpha=1$, there is no modification of the group velocity at leading order in $\mbA$ and hence no MDR contribution to $\Delta t$. Nevertheless, from Eq.\ (\ref{eq:DeltaPhi_vg}) we see there is still a constant $\Delta\Phi$. This appears due to the fact that the phase velocity of the wave does suffer corrections at linear order in $\mbA$ (see Eq.\ (\ref{vphase_linearA})). 
We emphasize that this constant phase shift in frequency-domain is important to take into account since it can also induce distortions in the GW waveform, analogous to what has been shown to happen for strong lensing of GWs \cite{Ezquiaga:2020gdt}. Further discussions and observational consequences of this constant phase shift will be addressed in Sec.\ \ref{sec:lensing}.

%------
%Particle velocity
%------
\subsection{Particle velocity} \label{sec:particle_velocity}

It has been common in the literature to interpret the GWs as a stream of particles that propagate according to the particle MDR that  the corresponds to the wave MDR of Eq.~(\ref{eq:MDR}):
\begin{equation} \label{eq:MDR_2}
    E^2 = c^2p^2\left[ 1 +\mbA (cp)^{\alpha-2} \right] \,, 
\end{equation}
with a correspondence between energy and frequency $E=\omega/a$ as well as momentum  and wavenumber $p=k/a$, which holds for the detected wavepacket at least locally around the detector.
In this interpretation each momentum component  propagates at the particle velocity $v_p$, defined by $v_p =c^2 p/E$. This set up was originally motivated by testing the graviton mass, where the particle and group velocity are the same \cite{Will:1997bb}, and later extended to Lorentz violating theories \cite{Mirshekari:2011yq}. Beyond a mass term, this approach is incompatible with the wavepacket propagation interpretation that underlies the WKB approach and its SPA decomposition.

For the dispersion relation under consideration, the particle velocity at linear order in $\mbA$ is:
\begin{equation}
    \frac{v_p}{c}\approx 1 - \frac{1}{2}\mbA\lp 2\pi f\rp^{\alpha-2}\,, 
\end{equation}
so that unlike for the group velocity, there is no value of $\alpha$ for which the correction vanishes at all frequencies as long as $\mbA\not=0$.

We follow the same procedure as in the previous subsection, and start by calculating the time delay between two different frequencies $f_s$ and $f_s'$, assuming they propagate at $v_p$. At leading order in $\mbA$, in agreement with \cite{Mirshekari:2011yq}, we obtain 
\begin{equation}
    \Delta t = (1+z)\left(\Delta t_s +\frac{\mbA}{2}D_\alpha\left[\frac{1}{(2\pi f_s)^{2-\alpha}}-\frac{1}{(2\pi f_s')^{ 2-\alpha}}\right]\right)\,.
    \label{eq:dt_vp}
\end{equation}
As before, we then obtain the frequency-domain phase by using Eq.\ (\ref{eq:fourier_phase}). For $\alpha\not=1$, we obtain:
\begin{align}\label{eq:psi_vp_not1}
    \Psi_j^{\alpha\not=1}(f)=&  2\pi\int_{f_{sj,c}}^{f_s}\lp t_{e,j}(f'_s)-t_{e,j}(f_{sj,c}) \rp df_s' -  \Phi_{j,c} - \frac{\pi}{4} 
     \\
    &  
     +
     2\pi f \left[ T_{j,c} - \frac{\mbA}{2}\frac{(1+z_s)D_\alpha}{(2\pi f_{sj,c})^{2-\alpha}}\right]  
     -\frac{1}{(1-\alpha)}\frac{\mbA}{2 }\frac{ D_\alpha}{(2\pi f_s)^{1-\alpha}}+
      \frac{(2-\alpha)}{(1-\alpha)}
      \frac{\mbA}{{2}}\frac{D_\alpha}{(2\pi f_{sj,c})^{1-\alpha}}   \,. \nonumber
\end{align}

Note that all the terms with explicit dependence on $\mbA$ differ from the group velocity result (\ref{eq:psi_tot}) by a constant factor $(1-\alpha)$. Consequently some terms diverge as $\alpha\rightarrow 1$, and for $\alpha=1$ Eq.\ (\ref{eq:psi_vp_not1}) does not apply. Instead, for $\alpha=1$ we obtain
\begin{align}\label{eq:psi_vp_1}
    \Psi_j^{\alpha=1}(f)=&      2\pi\int_{f_{sj,c}}^{f_s}\lp t_{e,j}(f'_s)-t_{e,j}(f_{sj,c}) \rp df_s'   - \Phi_{j,c}  - \frac{\pi}{4}
\nonumber\\
    & +
    2\pi f \left[ T_{j,c}-\frac{ \mbA}{2}\frac{(1+z_s)D_1}{2\pi f_{sj,c}}\right]  + 
    \frac{\mbA D_1}{2}\ln\left(\frac{ f_s}{f_{sj,c}}\right) +\frac{1}{2}D_1 \mbA  \,. 
\end{align}
This is  different from the group velocity result (\ref{eq:psi_tot}) not just in amplitude but in the logarithmic functional form.  

We can now reinterpret the various groups of terms in Eqs.\ (\ref{eq:psi_vp_not1}) and (\ref{eq:psi_vp_1}) as we did for the group velocity.
 It is again the case that the term in square brackets  vanishes because
 of the explicit form for the time delay of the coalescence frequency, 
 but now for signals traveling at the particle velocity.
 
Unlike the group velocity case where wave propagation links the emitted $\Phi_j^e$ and observed SPA phase 
$\Phi_j$ via Eq.\ (\ref{eq:DeltaPhi}), for this particle velocity case there is no specified prescription for connecting the phases in the literature.  However we can demand consistency in the frequency dependence of $\Phi_j(f)$.
In particular, if we again express the emitted and observed phase relation as: 
\begin{equation}
    \Phi_{j}(f) = \Phi^{e}_{j}(f) + \Delta\tilde\Phi (f) \,,
\end{equation}
from Eqs.\ (\ref{eq:Phij}) and (\ref{eq:dt_vp}) we can see that the MDR part of $\Phi_j$ scales with 
frequency as $\int  f^{\alpha-1}  d\ln f$.  For $\alpha> 1$, the MDR modification  and hence $\Delta\tilde\Phi(f)$ vanishes
at $f\rightarrow 0$ and for
$\alpha<1$ at $f\rightarrow \infty$.  Therefore:
 \begin{equation} \label{eq:Dphi_alpha_part}
\Delta \tilde\Phi(f) =
\begin{cases}
\dfrac{2-\alpha}{1-\alpha}\,\dfrac{\mbA }{2} \dfrac{D_\alpha}{(2\pi  f_s)^{1-\alpha}} & (\alpha\ne 1)\\
 -\frac{1}{2}  \mbA D_1  \ln\left(\dfrac{f}{f_0} \right)&
 (\alpha=1)
\end{cases} \quad .
\end{equation}
Here $f_{0}$ is the frequency at which 
$\Delta\tilde{\Phi}(f_0)=0$, and it is $f_0=\infty$ for $\alpha<1$,
$f_0 \rightarrow 0$ for $\alpha>1$ and
is unknown a priori for $\alpha=1$ unless one supplements the phenomenological MDR (\ref{eq:MDR_2}) with a physical theory connecting  particle propagation to wave propagation.

Notice also that unlike the group velocity, the seemingly obvious assumption that the phase of the signal propagates at the phase velocity is inconsistent:
\begin{align} 
 \int_{t_e}^{t} \left[ v_\text{ph}(t') - v_p(t')\right] p(t') dt' \label{eq:notdeltaphi_vp}
&=  \mbA \frac{D_\alpha}{(2\pi f)^{1-\alpha}} \ne \Delta\tilde \Phi(f)
\,,
\end{align} 
for any $\alpha\ne 0$. For example, for $\alpha=2$ this assumption would predict GW waveform distortions because the phase velocity differs from the particle velocity even though the MDR still looks like a frequency independent shift from the speed of light $c$. The apparent inconsistency of Eq.\ (\ref{eq:notdeltaphi_vp}) happens because the particle velocity case does not follow the physical principles of wave propagation and thus $v_{\rm ph}$ does not determine the phase evolution of the signal in this case.

Altogether, we thus obtain:
\begin{align}
    \Psi_j&\equiv \Psi_j^{\rm GR}(f)+\Delta\tilde{\Psi}(f) \label{eq:DeltaPsi_1}
\end{align}
where $\Psi_j^{\rm GR}(f)$ is given explicitly by Eq.\ (\ref{eq:Psi_GR}), and
\begin{align}
 \Delta\tilde\Psi(f)&=
 \begin{cases}
 - \dfrac{ \mbA }{2(1-\alpha)} \dfrac{D_\alpha}{(2\pi f_s)^{1-\alpha}}
 & \alpha \ne 1
 \\
  \frac{1}{2} \mbA D_1 \ln\left( \dfrac{e f }{f_0}\right) & \alpha=1
  \end{cases}
  \quad ,
  \label{eq:DPsitilde}
\end{align}
where $e$ is Euler's number. 
For $\alpha\neq1$ this expression is different to the SPA result in (\ref{Psi_spa}) by a factor $1/(1-\alpha)$, and therefore for $\alpha=0$ both approaches coincide. This happens because the particle and group velocities are the same for $\alpha=0$. 
On the other hand, 
for $\alpha=1$, the particle velocity approach leads to a logarithmic phase correction, with the frequency at which $\Delta\tilde \Phi=0$, $f= f_0$, differing from that of $\Delta\tilde\Psi=0$, $f=f_0/e$
due to the arrival time difference of the $f_0$ frequency.

Note that with this derivation we have demonstrated that for $\alpha\neq1$ the constraints from the group velocity can be easily translated to those from the particle velocity by using the aforementioned rescaling even in the presence of higher modes, generalizing~\cite{Abbott:2020jks}.  Furthermore,
most previous studies in the literature \cite{LIGOScientific:2019fpa,Abbott:2020jks,LIGOScientific:2021sio, Mastrogiovanni:2020gua, Baker:2022rhh} have analyzed the propagation of GW signals with only the $j=22$ mode, in which case any constant phase shift in $\Delta\tilde{\Psi}$ is completely degenerate with  the orbital coalescence phase of the binary $\varphi_c$, and have therefore removed any constants in $\Psi_{22}(f)$ associated with $\Phi_{22,c}$ as well as shifts in the arrival time of the coalescence frequency $T_{22,c}$.
For the same reason, previous studies have not made a distinction between $\Delta \tilde{\Psi}(f_{22,c})$ and $-\Delta \tilde{\Phi}(f_{22c})$, since
the two differ by these constants.
For $\alpha\ne 1$ we have shown that the result using the 22-degeneracy and dropping these distinctions for $\Delta\tilde\Psi$ is the same as retaining them, for any type of  mode, with  a specific choice for the global reference time $T=0$. 
For $\alpha=1$, the 22-degeneracy makes the choice of $f_0$ irrelevant for 22-waveform analyses whereas for non-degenerate higher modes it becomes relevant, albeit only weakly so if  no observed frequency is close to $f_0$, as we will confirm in the next section.  
The LVK collaboration has adopted the convention of choosing the normalization frequency to be inversely proportional to the chirp mass $f_0/e=c^3/\pi G\mathcal{M}_z$ of the binary system.  
Notice that this convention is unphysical when considering multiple sources since the reference frequency depends on the mass of the source and is not intrinsic to the modified theory of propagation.
Even though for GW signals with higher modes a constant phase shift in frequency-domain can introduce distortions to the extent that $f_0/f_{22,c}$ is still finite, in this paper we will make the same $f_0$ choice as LVK, unless specified otherwise.

%------
%Comparison propagation approach
%------
\subsection{Comparison of approaches}
We have shown that the WKB approach yields exactly the same result as the SPA approach using group velocity, for linear corrections beyond GR, since they both consistently rely on wave propagation in the MDR. We will henceforth call these results as group velocity results. However, we have also found that the SPA approach using particle velocity always yields a different result, albeit with universal frequency dependence for all modes,  for any $\alpha \ne 0$ of the MDR for the various phase shifts in the frequency-domain, which we will refer to as particle velocity results.
For the reader's convenience, we summarize the various phases that we have defined and the equations in which they are defined in Table  \ref{table:notation}. 

%TABLE FOR NOTATION
\begin{table}[h!]
\centering
\begin{tabular}{|c|c|m{10cm}|} 
 \hline
 Phase & Eq. &  Meaning \\ 
 \hline
 $\Phi_j(T)$ & (\ref{eq:Phi_j}) & Time-domain phase of a mode $j$ in the signal\\
 $\Psi_j(f)$ & (\ref{eq:Psi_j}) & Frequency-domain phase of a mode $j$ of the signal at $T=0$ \\
 $\Phi_{j,c}$ & (\ref{eq:Phi_j}) & Detected coalescence phase of mode $j$\\
 \hline
  $-\Delta \Phi(f)$ & (\ref{eq:DeltaPhi_vg})&  Shift from emitted to detected phase with $v_g$\\
 $\Delta\Psi(f)$ &
 (\ref{Psi_spa}) & Shift from GR to MDR at $T=0$, with $v_g$ \\
  $-\Delta\tilde\Phi(f)$ & (\ref{eq:Dphi_alpha_part}) & Shift from emitted to detected phase  with $v_p$ \\
 $\Delta\tilde\Psi(f)$ & (\ref{eq:DPsitilde}) & Shift from GR to MDR at $T=0$, with $v_p$\\
 \hline
\end{tabular}
\caption{Summary of phase definitions. Here, MDR refers to the modified dispersion relation that was parametrized as in Eq.\ (\ref{eq:MDR_2}). Also, $v_g$ and $v_p$ refer to the group and particle velocity, respectively. ``Shift" refers to the frequency-domain phase shift.
}
\label{table:notation}
\end{table}

%ALPHA = 0
In order to gain some intuition about the effect of the different MDRs on the detected signals, we study the time-domain waveforms of signals that include higher modes. For concreteness, in all cases, we will analyze the GW signal from a black hole binary with asymmetric masses, in a nearly-circular orbit, where the GW modes $j$ are given by different pairs of spherical harmonic numbers. 
In the plots of this section, we will choose a binary with redshifted chirp mass $\mathcal{M}_z=37.5M_\odot$ and mass ratio $q=0.1$.

\begin{figure*}[t!]
\centering
\includegraphics[width=0.95\textwidth]{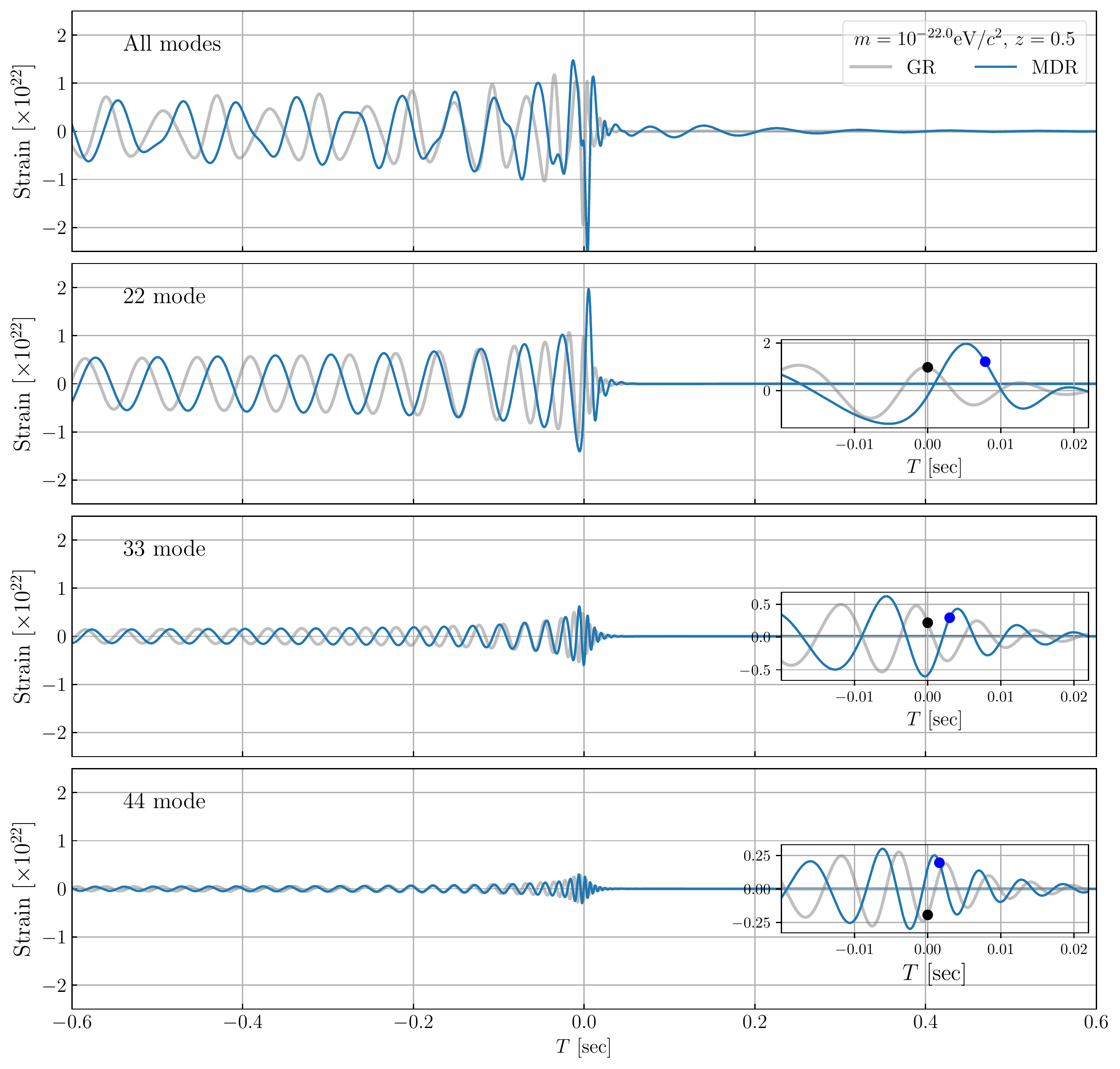}
\caption{\label{fig:prop_mass} 
Wave-distortions induced by an MDR with  $\omega^2 = c^2k^2 + m^2c^4$
, i.e.\ $\alpha=0$ in Eq.\ (\ref{eq:MDR}). 
Only in this case the modified dispersion relation (MDR) is the same for 
both the group and particle velocity results. 
We plot the total waveform and selected modes in the different panels. 
In the inset plots we highlight with a dot the arrival time of the coalescence frequencies for each mode, which were all emitted at a common time. 
$T=0$ corresponds to the arrival time of the coalescence frequencies in GR as given by the \texttt{IMRPhenomHM} \cite{PhysRevLett.120.161102} waveform approximant. 
The original (GR) waveform is an unequal mass binary with redshifted chirp mass $\mathcal{M}_z=37.5M_\odot$, mass ratio $q=0.1$, without spin, located at $z=0.5$, with inclination $\iota=\pi/3$, 
and Euler angles of the detector orientation with respect to the radiation frame $(\theta,\phi,\psi)=(0.3,0.4,1.5)$
 following the conventions of \cite{Ezquiaga:2020gdt}.
}
\end{figure*}

Let us start by considering the most studied case---the propagation of a massive graviton (corresponding to $\alpha=0$). 
This is the only case in which the group and particle velocity results are the same. In Fig.\ \ref{fig:prop_mass} we compare the GW signal from 
GR (grey line) versus the MDR signal  (blue line) during propagation, assuming a GR emission. In the top panel, we plot the total GW signal. In the lower panels, 
we plot the most relevant GW modes associated to the harmonic multipoles $j= \ell |m| =22, 33, 44$. As we can see, the higher the harmonic multipoles, the smaller the distortions. This is because higher GW modes carry higher frequencies, and for $\alpha=0$ the propagation speed of GWs deviates less from GR for higher frequencies. 
For $j=22$, this GW mode suffers from a squeezing compared to the GR signal because the early (lower) inspiral frequencies travel slower than the frequency components near the merger, which can then catch up. 
In addition, in the inset panels we highlight the part of the signal around $T=0$ for each mode (recall that $T=0$ is defined as a global time $t=t_c$ for all modes).
The dots indicate the arrival time of the $f_{j,c}$ frequencies for each mode, which were all emitted at the same time. 
The frequency of each mode can be computed from the time derivative of their phase following Eq.\ (\ref{eq:freq_from_phase}).
Here, we again explicitly see that the higher the mode, the shorter the associated time delay. 
This allows us to understand how in the total signal most of the distortions at $T<0$ come from the interference of the different modes, while the signal at $T>0$ is dominated by the 22-mode. 
In this example, the modification of the phase at the coalescence frequency for each of the modes is given by $\Delta\Phi(f_{22,c})=2.2\pi$, $\Delta\Phi(f_{33,c})=1.4\pi$ and $\Delta\Phi(f_{44,c})=\pi$. 
We can verify these phase shifts by comparing where the black and blue dots are located with respect to the peaks and troughs of the signal. 
From here we can see that $\Delta\Phi(f_{j,c})$ is a more direct observable compared to $\Delta\Psi(f_{j,c})$, which differs here by a factor of $-2$, recalling (\ref{eq:phaseratio}).

%---------
%FIGURE MASS TERM
%--------

%ALPHA = 1

%---------
%FIGURE GROUP VELOCITY V.S. PARTICLE VELOCITY
\begin{figure}[h!]
\centering
\includegraphics[width=0.95\textwidth]{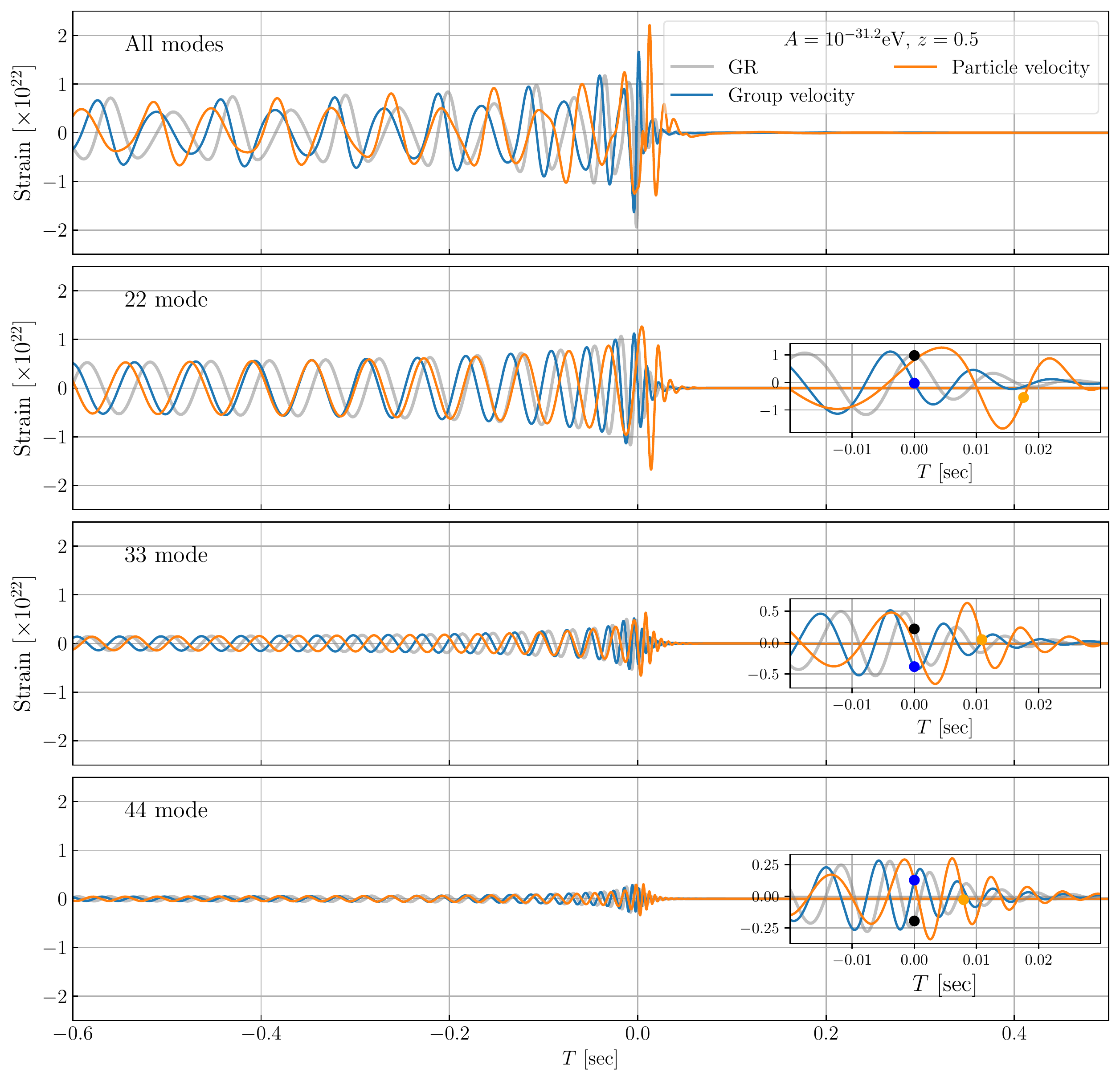}
\caption{ 
Wave-distortions induced by a modified GW propagation with dispersion relation $\omega^2=c^2 k^2 + \mbA c a k $, i.e. $\alpha=1$ in Eq.\ (\ref{eq:MDR}). 
We compare the stationary phase approximation   
result with the group velocity (blue lines), which is equal to the WKB approach, with the result assuming that GWs propagate at the particle velocity (orange lines). 
We plot the total waveform and selected modes in the different panels. 
In the inset plots we highlight with a dot the arrival time of the coalescence frequencies  for each mode. There is no time delay for the group velocity.  
We use the same source parameters and time conventions than Fig. \ref{fig:prop_mass}. 
We choose $f_0/e=c^3/\pi G \mathcal{M}_z$ as the normalization frequency for the particle velocity modification, see\ (\ref{eq:DPsitilde}), 
following LVK convention. 
}\label{fig:waveform_alpha_1}
\end{figure}
%---------

Next we consider the case of $\alpha=1$, that corresponds to a linear frequency correction to the GR dispersion relation. In this case, the group and particle results differ, so in each panel of Fig. \ref{fig:waveform_alpha_1} we plot three curves corresponding to GR (grey), MDR group velocity (blue), MDR particle velocity  (orange). 
We follow the same plotting conventions as Fig.\ \ref{fig:prop_mass}. 
For concreteness, we fix the normalization frequency of the particle velocity MDR to the LVK convention (see below (\ref{eq:DPsitilde})). 
The group velocity approach gives a frequency-independent phase correction to the waveform, which can be seen from the bottom three panels, where for each mode the peaks of the blue curve have a constant phase shift compared to those of the grey curve. In this example, this constant phase shift corresponds to $\Delta \Phi=2.5\pi$ and correspondingly shifts the phase of all modes by $\pi/2$, which we shall see below can mimic the effects of lensing.
This constant $\Delta \Phi$ brings non-trivial net distortions in the total waveform, when the modes are superimposed as can be seen from the top panel. 
Furthermore, the group velocity approach predicts $v_g=c$ for $\alpha=1$, and thus the group velocity waveform in blue has no time delay compared to GR. This can be seen explicitly in the insets of the bottom three panels, where we indicate in dots the arrival times of the coalescence frequencies $f_{j,c}$ for each mode. Here we can see that for GR (black dots) and MDR group velocity (blue dots), the frequencies  $f_{j,c}$ arrive at the same time for all modes.

On the other hand, the particle velocity approach leads to both a frequency-dependent phase correction to the waveform, in addition to a frequency-dependent time delay. More precisely, this approach predicts a logarithmic running phase shift in frequency-domain. 
This can be noticed most clearly in the (33) mode panel, where for the same value of $\mbA$ at the lowest frequencies the GR and particle velocity signals are nearly in phase for various cycles because during the inspiral the frequency evolution of the mode is very slow, and hence the logarithmic phase shift is nearly constant across these first cycles. However, the two waveforms start getting out of phase as we approach the merger, where the frequency evolution of $h$ is faster, finally arriving at
$\Delta \tilde \Phi_{22,c} \approx 13.3\pi $,
$\Delta\tilde \Phi_{33,c} \approx 12.1\pi$,
$\Delta\tilde \Phi_{44,c} \approx 11.4\pi$. 
The time delay effect is directly seen in the inset panels where the orange dot indicates the arrival time of $f_{j,c}$ for the particle velocity waveform, which always arrives after the black dots, and where the amount of time delay varies for the various modes. 
We can again see that $\Delta\tilde \Phi_{j,c}$ gives the phase change at that arrival time. 
Finally, notice that for the same value of $\mbA$, the particle velocity approach predicts a much larger $\Delta\tilde\Phi_{22,c}$ than the group velocity
$\Delta\Phi_{22,c}$ due to the LVK convention where there is a large ratio of   $f_0/f_{22,c}$.  Since $\Delta\tilde \Phi_{j,c}$ is the direct observable, we shall see in the next section that when fitting to 
observed waveforms it is the phase that is held fixed and so the best fit amplitude $\mbA$ is much smaller.   This also has the consequence of bringing the
particle velocity fits much closer to the group velocity fits by effectively reducing the amplitude of the logarithmic running.

%------------
%SECTION
%------------ 
\section{Degeneracies with strong lensing}\label{sec:lensing}

We have seen throughout the paper that a modified GW propagation generically changes the phases and arrival times of different frequency components.
In order to assess if such a modification of gravity is being detected it is crucial to understand if the same effect could be degenerate with any astrophysical or propagation effect within GR. 
As a concrete example we consider the possible degeneracies between a modified dispersion relation and strong lensing. 
As discussed in detail in \citep{Ezquiaga:2020gdt} (see also \cite{Dai:2017huk}), strong lensing not only induces magnifications and time delays of the detected signals, but it also changes the phase of the waveform. This phase effect always happens, even when the wavelength of the signal is much smaller than all the relevant lensing  scales (which corresponds to the regime that in the literature is usually treated using geometric optics). The images formed due to strong lensing in the geometric optics regime can be classified into three types: I, II, and III---distinguished by whether the stationary point they correspond to in the time delay surface is a minimum, saddle or maximum, which itself depends on the lens properties and source-lens-observer geometry. In the frequency-domain, any lensing signal $h_L$ can be expressed as a sum over the ``images" indexed by $j$: 
\begin{equation}
\label{eq:lensedwaveform}
    h_L(f)= h(f)\sum_j \vert\mu_j\vert^{1/2}\exp\left(i 2\pi f t_{d,j}-i\, n_j\pi\right) ,
\end{equation}
where $f>0$ is the frequency, $\mu_{j}$ is the magnification, $t_{d,j}$ is the time delay, and $n_j$ takes the values $0, 1/2, 1$ for type I, II, and III images, respectively. Here, $h(f)$ is the signal that would arrive without lensing, assumed to be given by GR in this paper. If the separation in time delays is longer than the signal in the detector, then we can consider the waveform of each image separately.
Note that in complete analogy with the $\alpha=1$ group velocity case, the constant phase shift introduced by strong lensing can distort the time-domain waveform in a manner that cannot be mimicked by changing source parameters if higher modes are present, as we shall see below. 
In turn, this means that strong lensing could be identified with a single image \citep{Ezquiaga:2020gdt,Wang:2021kzt,Janquart:2021nus,Vijaykumar:2022dlp}. This phase information is also useful in multiple image searches to determine the image types and constrain the lens model \cite{Dai:2020tpj,Liu:2020par,LIGOScientific:2021izm}, being already implemented in different pipelines \cite{Lo:2021nae,Janquart:2021qov}.

In this section we will make an exploration of the degeneracies between different MDRs and strong lensing.  
More concretely, we will consider 
type II strongly-lensed GR signals that exhibit multiple modes, and compare the goodness of fit of unlensed GR templates versus MDRs templates. The detailed methodology applied in this paper is explained in Subsec.\ \ref{sec:chi_method}, while numerical results for various examples are discussed in Subsec.\ \ref{sec:prop_vs_lensing}. These results will illustrate how some strongly-lensed GW events in GR can misleadingly show preference for modified gravity over unlensed GR, and emphasize the need for adding astrophysical propagation effects into considerations in templates in future data analyses. 
As a comparison, we will also analyze the LVK phenomenological parametrization, which follows from particle velocity approach  presented in Sec.\ \ref{sec:particle_velocity} rather than the wave propagation or group velocity approach.

We focus our analysis on non-spinning sources with asymmetric masses computed using \texttt{IMRPhenomHM} \cite{PhysRevLett.120.161102} waveform approximant.

\subsection{Methodology}\label{sec:chi_method}

In order to analyze the approximate degeneracies between GR and modified gravity parameters we follow the standard approach (see e.g.\ \cite{Maggiore:1900zz}) of quantifying the matching between a given detected signal and different templates, in the presence of measurement noise of a single detector. In order to do this we introduce the inner product of two functions $a$ and $b$ in the frequency-domain:
\begin{equation}
    (a\vert b)=4\, \text{Re}\left[\int_0^\infty df \frac{\tilde{a}(f)\cdot \tilde{b}^*(f)}{S_n(f)}\right]\,,
\end{equation}
where $S_n(f)$ is the one-sided noise power spectral density. 

For a measurement of the strain $s(t) = h(t)+n(t)$ consisting of a GW waveform $h(t)$ and noise $n(t)$, 
the signal-to-noise ratio (SNR) is defined as
\begin{equation}
    \rho = \sqrt{( h\vert h )}\,.
\end{equation}
Then the mismatch between a given template $h_T$ and a signal can be quantified as 
\begin{equation}
\label{eq:mismatch}
    \epsilon =  1-\frac{(h|h_T)}{\sqrt{(h|h)(h_T|h_T)}}\,.
\end{equation}
 
In the context of GW parameter estimation the log-likelihood of a signal $s$ given Gaussian noise and a waveform model with a set of parameters $\theta_T$ with $h_T = h(\theta_T)$ is determined by:
\begin{equation}
    \ln \mathcal{L}= -\frac{1}{2}(s-h(\theta_T)|s-h(\theta_T))+C\,,
\end{equation}
where $C$ is a normalization constant. One way of measuring how well the morphology of the signal matches the template is through the $\chi^2$ goodness-of-fit test. It is defined from the exponent of the likelihood, therefore just being
\begin{equation}
 \chi^2=(s-h_T|s-h_T)=(h|h)-2(h_T|h)+(h_T|h_T) + (n|n)\,,   
\end{equation}
where in the second equality we have expanded the inner product and averaged over noise realizations  assuming that the noise is uncorrelated with the signal and template.

Our goal is to compare the fit of a given template $h_T$ to the true signal. The increase in $\chi^2$ of the template with respect to the truth is 
\begin{equation}
\begin{split}
 \Delta\chi^2&=\chi^2_\text{template}-\chi^2_\text{truth}=(h|h)-2(h_T|h)+(h_T|h_T)\,. 
\end{split}
\end{equation}
Note that in the GW literature the following notation is also common $\Delta\chi^2 \equiv ( \delta h| \delta h)$, where $\delta h = h-h_T$ \cite{Lindblom:2008cm}.
If we use different templates and compare their $\Delta\chi^2$, this difference will tell us how much better one signal fits one template over the other one.

We can rewrite this expression in terms of $\epsilon$:
\begin{equation}
    \Delta\chi^2=(h|h)-2(1-\epsilon)\sqrt{(h_T|h_T)(h|h)}+(h_T|h_T)\,.
\end{equation}
Now if the difference between the templates and the signal is small, $\rho^2\equiv (h|h)\approx (h_T|h_T)$, the above equation simplifies to 
\begin{equation}
    \Delta \chi^2 \simeq 2\rho^2\cdot \epsilon\,.
\end{equation}
The advantage of this expression is that $\epsilon$ is essentially only a statement of the morphology of the signals and, therefore, one can scale it to the SNR of the event in the detector to determine if a modification could be detected. 
In the frequentist interpretation, an improvement of $\Delta\chi^2= X^2$ from minimizing over a single parameter corresponds to an $X \sigma$ preference for adding that parameter.   Note then that an improvement in $\Delta\chi^2/\rho^2 = Y^2 \approx 2\epsilon$ corresponds to $(Y\rho) \approx \sqrt{2\epsilon}\rho$ in units of
$\sigma$.
In other words, with a high signal to noise event, even a fractionally small adjustment in the template can give a significant preference for an additional parameter.  In a Bayesian framework, $\Delta\chi^2$ provides the likelihood ratio for the improvement in the posterior probability which in general is  then weighted with the prior, and as such,  it can be interpreted as parameter constraints for a flat prior.

In the next subsection, we will quantify how well two templates---given by unlensed GR and an MDR---match  a strongly-lensed type II image, by obtaining the best-fit parameters of the templates and their associated $\chi^2$ fits. 
In principle, this analysis would require us to vary all the parameters in the templates, such as intrinsic source parameters (e.g.\ masses and spins), extrinsic source parameters (e.g.\ inclination, coalescence phase, merger time, distance)
and detector Euler angles  with respect to the radiation frame $(\theta,\phi,\psi)$,
 as well as modified gravity parameters. In practice, in this paper we take a simplified approach in which we only vary parameters that are theoretically expected to exhibit considerable degeneracy with strong lensing effects which cannot be broken with multiple detectors.  For example there is an orientation degeneracy that can mimic a constant phase shift for $|m|=\ell$ modes in a single detector, but this shift would be different for each detector \cite{Ezquiaga:2020gdt}. 
 Assuming that the degeneracies with the Euler angles can be broken, we have checked that the specific choice of
 values for $(\theta,\phi,\psi)$ has a negligible effect on $\Delta\chi^2/\rho^2$ and, as a consequence, we will leave these angles fixed throughout the analysis.

More specifically, we will only vary the coalescence phase $\varphi_c$, the two parameters $\mbA$ and $\alpha$ in the modified dispersion relation, and the  arrival time shift 
$T_{\rm shift}$, while keeping all the other parameters fixed to be the same in the signal and the templates. 
The coalescence phase is a special parameter as it is the only one that, when varied, does not change the polarization, the frequency evolution, nor the higher mode structure of a GW signal, analogous to the strong lensing phase shift. 
In GR, the relation between $\varphi_c$ and $\Phi$ is known. In fact, when $\varphi_c$ is varied, then the emitted phase $\Phi^e$ of each harmonic of order $|m|$  varies by a constant 
\begin{equation}\label{eq:varphic}
\Delta \Phi^e= |m|\Delta\varphi_c.
\end{equation}
Therefore given a signal with a dominant $m$ mode, $\varphi_c$ will be at least partially degenerate with the lensing phase shift in Eq.~(\ref{eq:lensedwaveform}) and the extent to which multiple  $|m|$ modes are detectable, the degeneracy is broken.
In addition, we will vary the merger time in the MDR templates because MDRs can lead to time delays and phase shifts that change the arrival time of the peak of the signal. 
Specifically, we vary $\varphi_c$ directly in the generation of the waveform with \texttt{pyCBC} \cite{alex_nitz_2019_3546372} and include a time shift $T_\text{shift}$ in the frequency domain. Altogether our frequency-domain templates will have as free parameters $\{\varphi_c,T_\text{shift},\mbA,\alpha\}$ and will be constructed as
\begin{equation}
    h_T(f|\varphi_c,T_\text{shift},\mbA,\alpha)=h_{\rm GR}(f,\varphi_c)e^{i\Delta\Psi(f,\mbA,\alpha)}e^{-i2\pi f T_\text{shift}}\,.
\end{equation}
Since the dimensions of the parameter $\mbA$ change for different values of $\alpha$, it will be convenient to renormalize it to a dimensionless quantity.  We choose this normalization to be given by the modified phase at the coalescence frequency, $\Delta\Phi( f_{22,c}) \propto \mbA$ (see Eq.\ (\ref{eq:DeltaPhi_vg}) and analogously
Eq.~(\ref{eq:Dphi_alpha_part}) for the particle velocity version $\Delta\tilde\Phi(f_{22,c}) \propto \mbA$).

Finally, let us point out that even though strong lensing changes the overall amplitude of the detected signal, here we do not vary the distance parameter of the templates as the main goal is to quantify the fits in the phase evolution of the waveform, regardless of how far the source is, and what magnification it has. We will ignore these amplitude effects, which change the total SNR by quoting  $\Delta\chi^2/\rho^2$. 
Similarly, the overall arrival time of the lensed signal $h_L$ will not be relevant, provided that we are in the regime of separate images. For these reasons the lens mass and location will not be relevant in our analysis. 
In any case, we choose a lens mass that makes the type II image detectable ($\rho\gtrsim 8$), and fix the detector sensitivity to A+ \cite{Aasi:2013wya}, expected for the fifth observing run (O5) in 2025+.
Throughout the analysis we will assume that the adopted waveform model is a true representation of the emitted signal. In reality, this is just a phenomenological model that carries some systematics that will become relevant in the high signal-to-noise limit, $\rho\gtrsim100$, see e.g. \cite{LIGOScientific:2016ebw}.

\subsection{Results}\label{sec:prop_vs_lensing}

%GR discussion
Let us suppose that a type II lensed GW signal is detected but not identified as such. 
The first question is how well an unlensed GR waveform would fit it. 
This question has been thoroughly studied in \cite{Ezquiaga:2020gdt} for different sources, and in \cite{Janquart:2021nus,Vijaykumar:2022dlp} and \cite{Wang:2021kzt} in the context of current and future GW detectors respectively.  In particular \cite{Ezquiaga:2020gdt} quantified the template 
mismatch parameter $\epsilon$ in Eq.~(\ref{eq:mismatch}) for unequal mass sources, spin, eccentricity and precession, which then determines the minimal SNR for which the unlensed GR waveform is a poor fit.

Unequal mass cases are interesting in that the presence of higher modes $|m|>2$ break the degeneracy between $\varphi_c$ and the frequency independent phase shift of $
\pi/2$ due to lensing as discussed around Eq.~(\ref{eq:varphic}).   The emission strength of
a $\ell\ell$ mode vs the 22 mode scales with the inclination $\iota$ of the source as 
$\sin^{\ell-2}\iota$. Hence as
 $\iota\rightarrow 0,\pi$, 
the $|m|=\ell>2$ higher modes are suppressed and in fact at those inclinations the signal has only
 $|m|=2$ for any $\ell$ due to the spin-2 nature of gravitational waves.  Therefore, in these cases, the $\varphi_c$ degeneracy is preserved and $\Delta\chi^2/\rho^2=0$. For this reason we choose the inclination to be  $\iota=\pi/3$, which is the median inclination of randomly oriented systems, with other values scaling roughly as
 $\Delta\chi^2/\rho^2 \propto \sin^2\iota$ for most cases of interest.
 The 50\% of the sources with inclinations larger than our median example
 have $3/4 < \sin^2\iota \le 1$, and so our example is representative of them as well, given this rough scaling.
 
\begin{figure}[t!]
\centering
\includegraphics[width=0.7\textwidth]{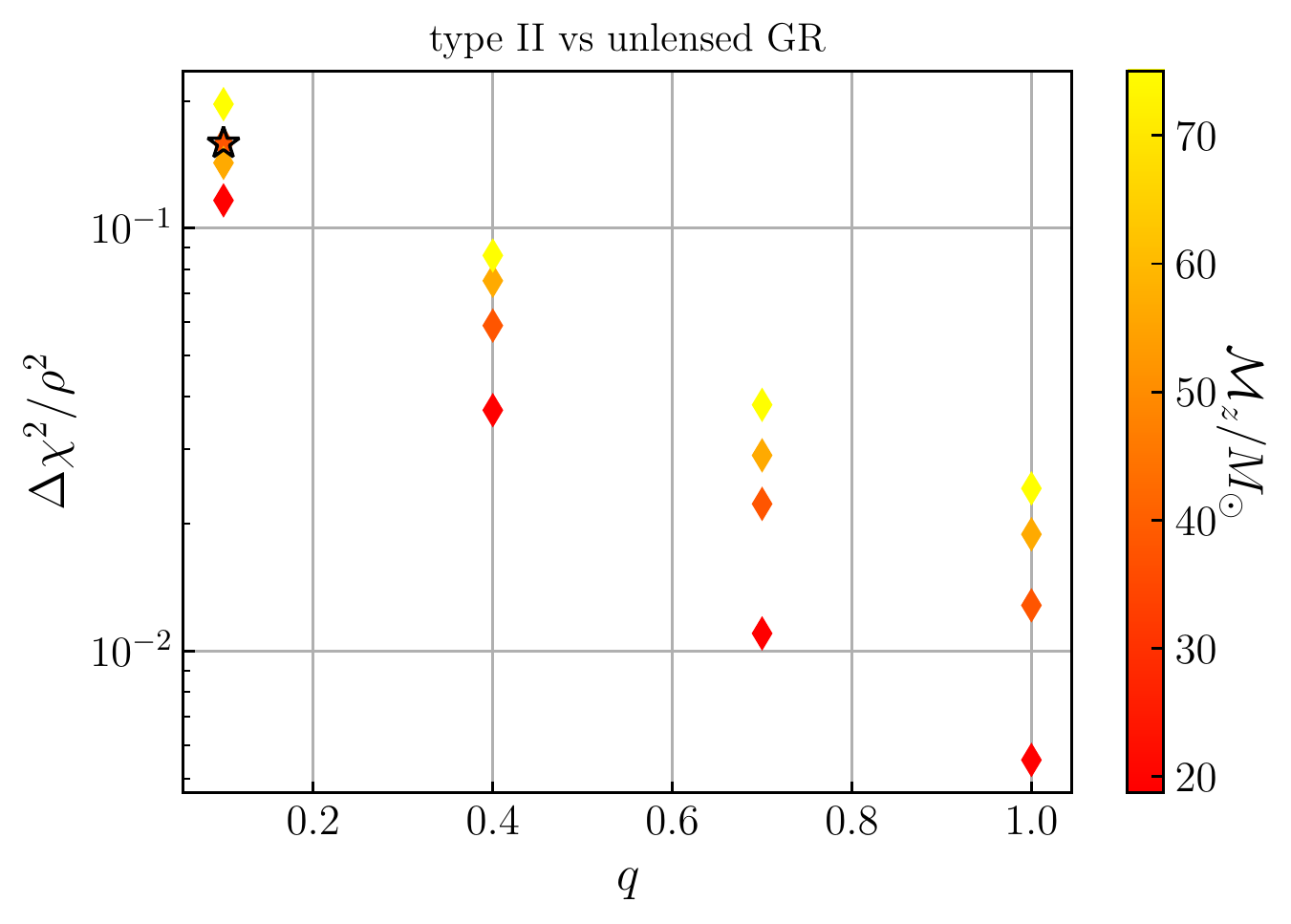}
\caption{\label{fig:chi2-q}
Goodness of fit $\Delta\chi^2$ of a lensed type II image with an unlensed GR template  as a function of the mass ratio $q=m_2/m_1$ for different values of the redshifted chirp mass $\mathcal{M}_z$.   Results are scaled by the squared SNR $\rho^2$, since $\Delta\chi^2\propto \rho^2$ for a fixed source and detector. 
For the GR template we minimize the $\Delta\chi^2$ varying both the coalescence phase and time. 
Our fiducial model, $\mathcal{M}_z=37.5M_\odot$, $q=0.1$ and $z=0.5$, is marked with a star, whereas the diamonds correspond to other cases with $\mathcal{M}_z=\{18.75,37.5,56.25,75\}M_\odot$.  
In all cases the inclination is fixed to $\iota=\pi/3$.
}
\end{figure}

In Fig.\ \ref{fig:chi2-q} we show the
mismatch of the unlensed GR template $\Delta\chi^2/\rho^2$ as a function of mass ratio $q$ after optimization over $\varphi_c$ and $T_{\rm shift}$ for a range of redshifted chirp masses. 
For definiteness we fix the orientation of the detector to $(\theta,\phi,\psi)=(0.3,0.4,1.5)$, whose variation changes the absolute but not the relative amplitude and phase of $|m|=\ell$ modes
\cite{Ezquiaga:2020gdt}.  So long as these orientation parameters are kept the same in the templates, the choice itself largely does not affect
 $\Delta\chi^2/\rho^2$.
As anticipated, for any given chirp mass, a better fit is achieved for more symmetric masses since the higher modes are emitted with a lower amplitude relative to the 22 mode. 

%--FIG: chi2 vs q
%----

The values of $\Delta\chi^2/\rho^2$ shown in Fig.\ \ref{fig:chi2-q} serve as a reference to understand for what kind of sources an unlensed GR template would provide a bad fit and be potentially mistaken for an MDR signal. 
For example, given a detection threshold of $\rho>8$, a typical event at this threshold SNR  would have a nominal ``$>3 \sigma$" bad fit of $\Delta \chi^2>9$  (i.e.\ $\Delta \chi^2/\rho^2 > 9/64$) whenever $q\lesssim 0.1$ and $\iota\gtrsim0.3\pi$.   
Of course for the rarer high SNR events that are well above detection threshold, a smaller $\Delta\chi^2$ mismatch at higher $q$ can be of equal significance.  
We emphasize that even for $q=1$ the fit is not exactly perfect since inclined sources still emit GW harmonics other than $22$.
In addition, we note that $\Delta\chi^2/\rho^2$ depends on $\mathcal{M}_z$ at fixed $q$ in a manner that reflects the relative SNR of higher modes for the given detector noise curve.
Indeed, for larger $q$, in the mass range shown and for the same SNR, the low frequency cut off of the detector reduces the relative contribution of the 22 mode 
during the inspiral phase, and  hence the relative importance of the higher frequency modes is larger as the mass increases.  
For very low $q\lesssim 0.1$ case, we instead find that the higher modes are so important that the trend with mass reverses since lower masses have a larger range of well-measured frequencies for the higher-mode mismatch to increase $\Delta\chi^2/\rho^2$. The combination of these two trends make the fit relatively insensitive to mass near $q=0.1$.

\begin{figure}[h!]
\centering
\includegraphics[width=\textwidth]{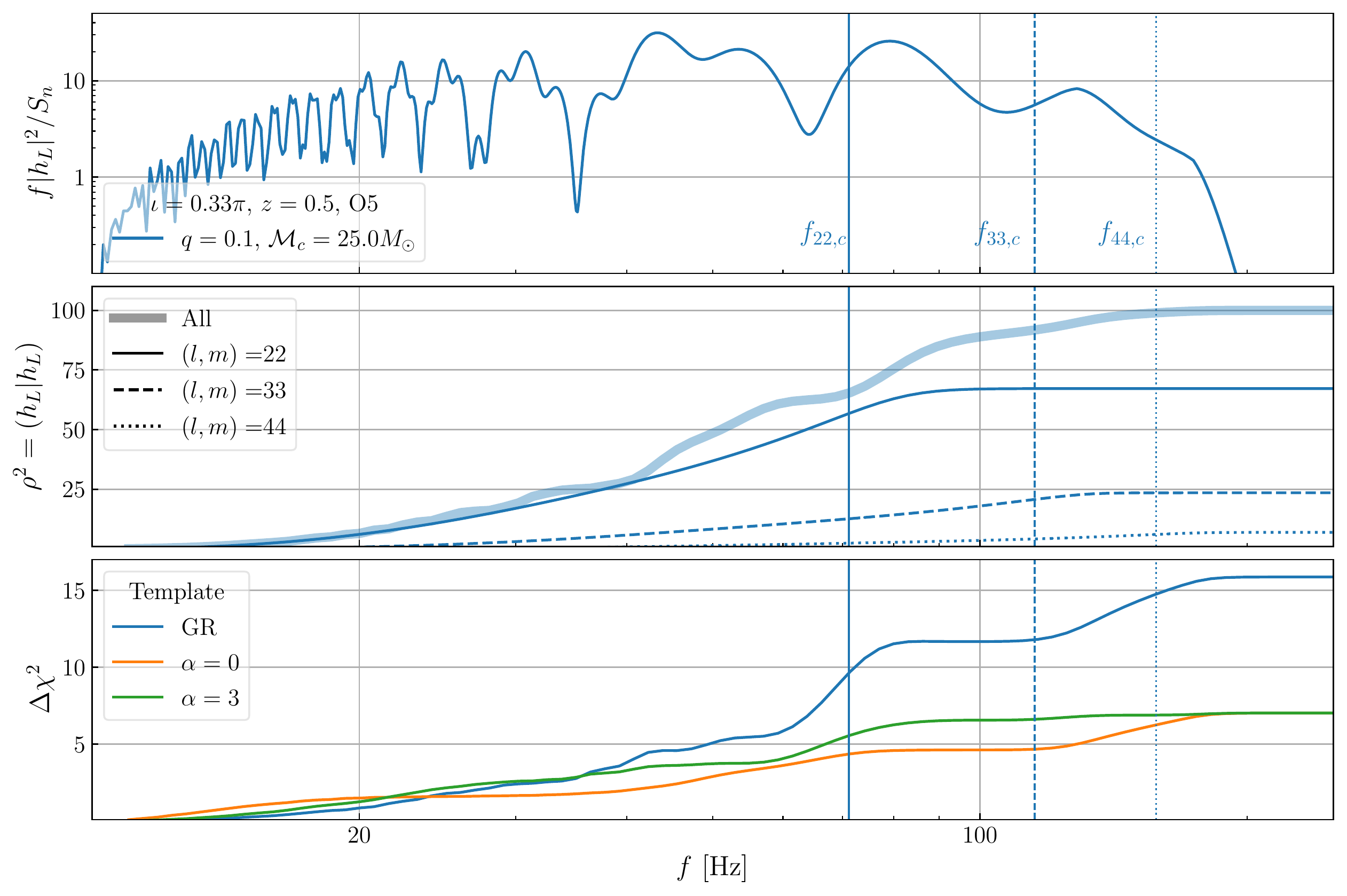}
\caption{\label{fig:chi2-f_lensing}
Lensed type II image signal in frequency domain and its fitting with different templates. The top panel shows the ratio of signal power and noise power per logarithmic frequency interval. The middle panel shows the cumulative squared SNR $\rho^2$ as a function of the maximum frequency for the total signal and the separate contribution of most dominant individual modes.
The bottom panel shows cumulative  $\Delta \chi^2$ when fitting the signal with an unlensed GR template, and with a template with a modified dispersion relation with $\alpha=0,3$. Vertical lines display the coalescence frequency of each mode. 
}
\end{figure}

Next, let us consider the case of $q=0.1$, $\mathcal{M}_z= {37.5} M_\odot$ more closely and analyze in detail how the unlensed GR template tries to fit the higher modes in the lensed signal. 
In the top panel of Fig.\ \ref{fig:chi2-f_lensing} we plot the signal power  $|h_L|^2$ 
 weighted by the noise power  per logarithmic interval in frequency.
 Noticeably the signal to noise diminishes at low frequencies $f\lesssim20$Hz due to the detector noise; the oscillations at intermediate frequency correspond to the constructive and destructive interference of the higher modes in frequency space; and
the sharp cut at high frequencies is due to the end of the signal after coalescence for this source. 

The role of higher modes becomes apparent in the middle panel of Fig.\ \ref{fig:chi2-f_lensing}  where we plot the cumulative  $\rho^2=(h_L|h_L)$ calculated between $f_\text{min}=10$Hz and $f$. 
We present this quantity for the total signal---which reaches a maximum SNR of  $\rho\simeq10$---as well as that of the three largest modes considered separately. It is clear here that the signal is dominated by the 22 mode, although 33 has an important  contribution and 44 has a small impact.  Other modes are
present but completely subdominant.
The SNR of the total signal has a ``step-like'' shape, accounting for the dominance of each higher mode as the frequency increases beyond the coalescence frequency of the 22 mode (see vertical lines indicating the coalescence frequency of the three harmonics). 
Furthermore, here we can see that $\rho^2$ reaches half  of its maximum well before the first merger frequency $f_{22,c}$, which means that, for this kind of source, most of the SNR comes from the inspiral regime. The template must therefore match a wide range of frequencies in the signal.  The cumulative $\Delta\chi^2$ plot in the bottom panel of  Fig.\ \ref{fig:chi2-f_lensing} helps us confirm this frequency-dependent behavior, where we see that the unlensed GR template fits well the lower half of frequency space in the cumulative SNR, where the inspiral 22 component dominates.  The template becomes a poor fit in the upper half of frequency space where the higher modes, especially 33, contribute more to the SNR.  
For the GR unlensed template, adjusting $T_\mathrm{shift}$ can bring the fit to higher modes into better alignment, but only for a small frequency range since the phase depends on 
$f T_\mathrm{shift}$ and there is no true time shift of the signal by design since we ignore the lensing time delay.   
In the top row of Tab.\ \ref{tab:chi2_baseline} we quote the best-fit parameters $\{\Delta\varphi_c,T_\text{shift}\}$ of this unlensed GR template---other parameters are relevant only to the MDR cases and are zero by definition here. 
From this table we can notice also that for a lower chirp mass (see $\mathcal{M}_z=18.75M_\odot$ model), the $T_\mathrm{shift}$ best-fit becomes smaller and $2\Delta\varphi_c$ approaches $\pi/2$ to mimic the lensing phase. This happens due to the fact that the 22 mode contains a higher proportion of the total SNR because of the extended frequency range of the detected inspiral. As a result,  $\Delta\chi^2/\rho^2$ also decreases. 
We emphasize that the presence of higher modes is crucial for the mismatch, since if the signal contained only $|m|=2$ modes, a template with $2\Delta\varphi_c = \pi/2$ and $T_\mathrm{shift}=0$ would produce a perfect fit of the signal.

%--FIG: chi2 vs freq for lensing
%----

\begin{figure*}[h!]
\centering
\includegraphics[width=0.49\textwidth]{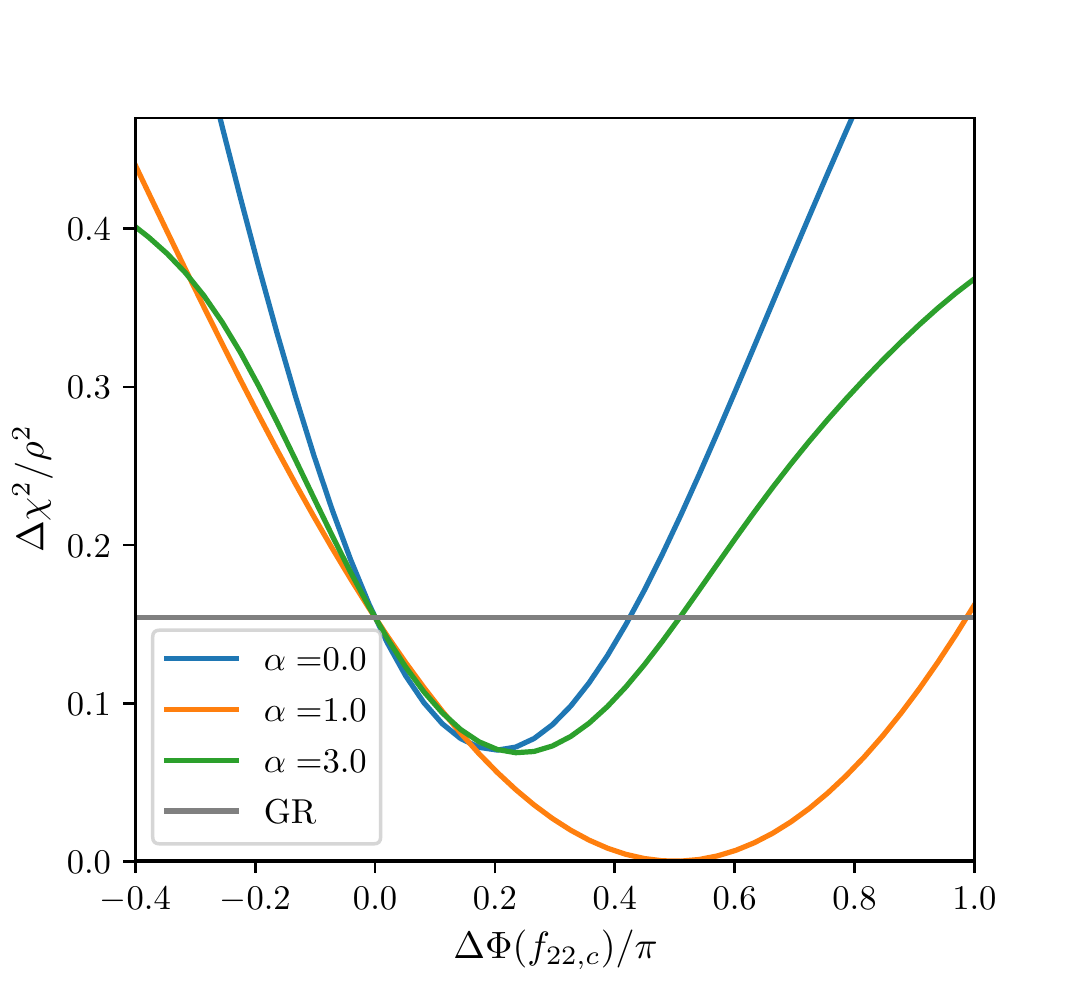}
\includegraphics[width=0.49\textwidth]{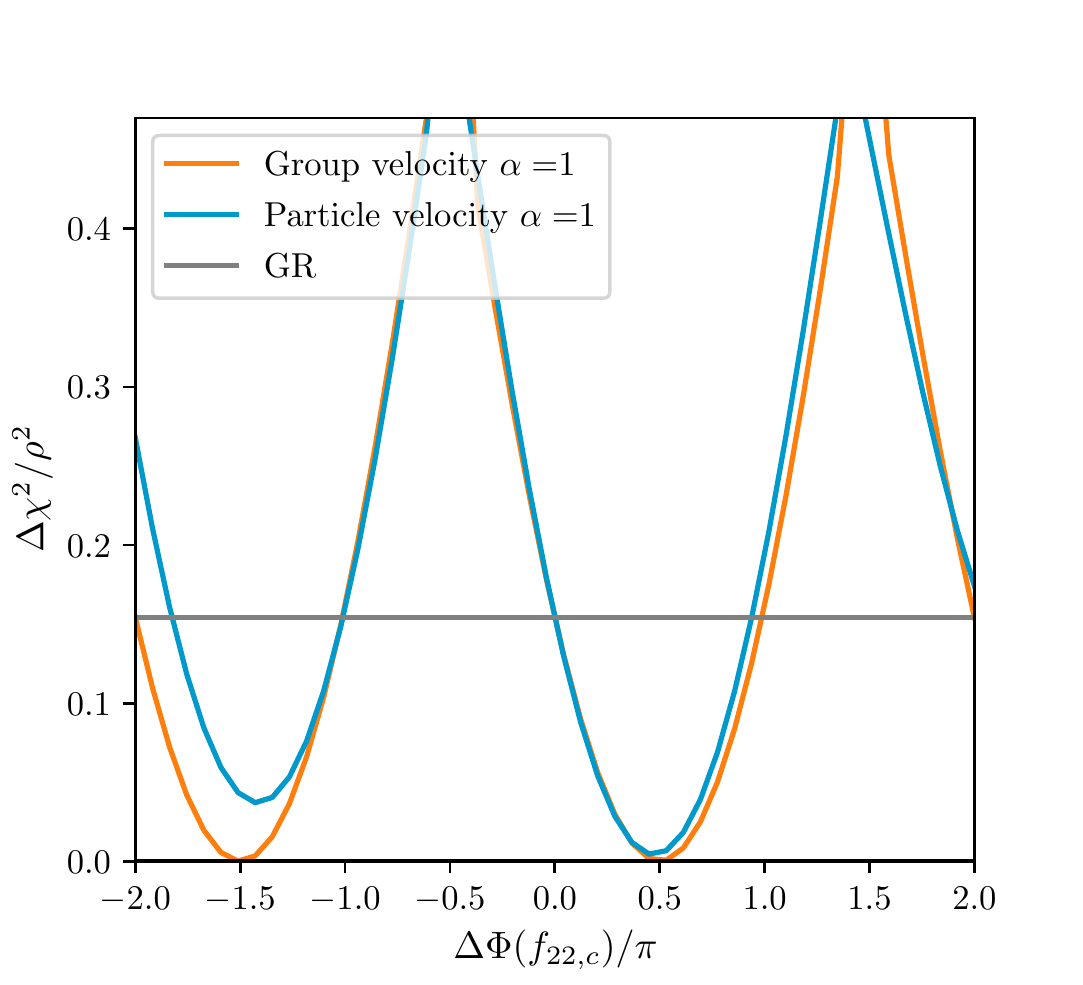}
\caption{\label{fig:chi2-A} 
Goodness of fit of a lensed type II image with a modified gravity template. 
We plot $\Delta\chi^2/\rho^2$ as function of $\Delta\Phi(f_c)$ for different MDRs. 
The change of the phase is proportional to the amplitude of the modified propagation, i.e.\ $\Delta\Phi(f_{22,c})\propto\mbA$. 
{\it Left:} we compare different $\alpha$'s with the group velocity approach. {\it Right:} we compare group velocity and particle velocity approaches with $\alpha=1$, fixing $f_0$ to the LVK convention.
For each $\Delta\Phi(f_{22,c})$, the coalescence phase $\varphi_c$ and $T_{\rm shift}$ are optimized to minimize $\Delta\chi^2$. We consider our fiducial binary with $\mathcal{M}_z= {37.5} M_\odot$.  
}
\end{figure*}

Having analyzed the unlensed GR template, the next question is whether an MDR template fits the signal significantly better than the unlensed GR template.
For concreteness, we focus on $\alpha=\{0,0.5,1,1.5,2.5,3\}$ motivated by modified gravity theories in the literature that predict $\alpha\geq 0$ in Tab.\ \ref{tab:chi2_baseline} where 
one additional parameter with respect to GR, $\Delta \Phi(f_{22,c})$, or
equivalently  $\mbA$ the MDR amplitude, is varied. Note that we are not quoting directly $\alpha=2$ since this is equivalent to the GR fit with $T_\text{shift}$.
In Fig.\ \ref{fig:chi2-A} we show how the $\Delta\chi^2/\rho^2$ depends on this parameter.  Each point is optimized over  $\varphi_c$ and $T_\text{shift}$ as in the GR case.  The extent to which the minimum lies below the unlensed GR line quantifies the improvement in the fit when optimizing the MDR amplitude.  The minimum values are also given in Tab.\ \ref{tab:chi2_baseline}. 
Note the best-fit phase $\Delta\Phi$ at $f_{22,c}$ depends on $\alpha$, which as we shall see below, reflects the fact that the best frequency for which to fit the lensing phase shift depends on the MDR. Nonetheless for each $\alpha$,
$\Delta\Phi(f_{22,c})\propto \mbA$ and
 to the extent to which $\Delta\chi^2/\rho^2$ is quadratic around the minimum our ``$3\sigma$" criteria on that quantity corresponds to an actual $3\sigma$ detection of a non-zero MDR amplitude parameter $\mbA$ or more generally a change of $\Delta\chi^2=1$ from the minimum gives its parameter errors for a flat prior.

%--FIG: chi2 vs DPhi

\begin{table}[h!]%[t]
    \centering
    \footnotesize
    \begin{tabular}{|c|c||c|c|c||c|c|}
    \hline
    \vphantom{$M_{z_z}^{z^z}$}
    Model ($18.75M_\odot$) & $\Delta\chi^2/\rho^2$ & $\Delta\Phi(f_{22,c})/\pi$ 
    & $\Delta\varphi_c/\pi$ & $f_{22,c} T_{\rm shift}$ & $\mbA$ [$10^{-15}\text{peV}^{2-\alpha}$] & $\Delta\Phi(f_{\rm eff})/\pi$ \\
    \hline
    Unlensed GR & 0.12 & 0 & 0.22 & -0.016 & 0 & 0\\
    \hline
    $\alpha=0$   & 0.081 & 0.072 & 0.14 & 0.064 & 3.28 & 0.14 \\
    $\alpha=0.5$ & 0.036 & 0.25 & 0.048 & 0.10  & 7.18 & 0.33 \\
    $\alpha=1$   & 0 & 0.50 & 0 & 0             & 10.1 & 0.50 \\
    $\alpha=1.5$ & 0.025 & 0.49 & 0.067 & -0.31 & 9.42 & 0.39 \\
    $\alpha=2.5$ & 0.068 & 0.30 & 0.16 & 0.38  & -1.26 & 0.18 \\
    $\alpha=3$   & 0.076 & 0.24 & 0.18 & 0.18  & -0.231 & 0.14 \\
    \hline
    $\alpha=1$(part.) & 0.0035 & 0.42 & 0.0014 & 0.083 & 2.02 & 0.47 \\
    \hline
        \hline \vphantom{$M_{z_z}^{z^z}$}
    Model ($37.5M_\odot$) & $\Delta\chi^2/\rho^2$ & $\Delta\Phi(f_{22,c})/\pi$ 
    & $\Delta\varphi_c/\pi$ & $f_{22,c} T_{\rm shift}$ & $\mbA$ [$10^{-15}\text{peV}^{2-\alpha}$] & $\Delta\Phi(f_{\rm eff})/\pi$ \\
     \hline
    Unlensed GR & 0.15 & 0 & 0.19 & 0.025 & 0 & 0 \\
    \hline
    $\alpha=0$   & 0.070 & 0.20 & 0.06 & 0.11    & 1.15 & 0.30 \\
    $\alpha=0.5$ & 0.022 & 0.39 & 0.0067 & 0.097 & 1.95 & 0.45 \\
    $\alpha=1$   & 0 & 0.50 & 0 & 0              & 2.53 & 0.50 \\
    $\alpha=1.5$ & 0.014 & 0.47 & 0.036 & -0.25  & 3.20 & 0.44 \\
    $\alpha=2.5$ & 0.055 & 0.31 & 0.10 & 0.47   & -0.924 & 0.32 \\
    $\alpha=3$   & 0.069 & 0.24 & 0.12 & 0.26   & -0.236 & 0.28 \\
    \hline
    $\alpha=1$(part.) & 0.0042 & 0.47 & -0.0055 & 0.069 & 0.570 & 0.49 \\
    \hline
    \end{tabular}
    \caption{\label{tab:chi2_baseline}
    Best fits to a type II strongly-lensed signal using unlensed GR templates and MDR templates.
       We optimize $\Delta\Phi(f_{22,c})$, $\varphi_c$, and $T_{\rm shift}$ simultaneously. 
    We consider sources fixed at $z=0.5$ with redshifted chirp masses $\mathcal{M}_z=18.75$ and $37.5 M_\odot$, and mass ratio $q=m_2/m_1=0.1$.
    The first row for each source shows the unlensed GR fit, the next six rows show MDR wave propagation fits for various $\alpha$, and the last row shows the case of $\alpha=1$ using the particle velocity approach using the LVK convention for $f_0$ and listing $\Delta\tilde\Phi$ in the relevant
     columns.  Other $\alpha$ values for the particle velocity fits simply rescale $\mbA$ by $1/(1-\alpha)$.  
    We show the inferred value of $\mbA$ with energy dimensions
      of picoelectronvolts (peV).  
    For the GR signal the SNRs are $\rho=7.1$ and $10$, and 
 $1/f_{22,c}=7$ and $14\,$ms, for the light and heavy sources respectively. 
   The last column presents the arrival phase at the frequency in which the template attempts to fit the signal. In a good fit this phase satisfies Eq.\ (\ref{feff_fit}).
    }
\end{table}

We have seen that strong lensing introduces a frequency-independent phase modification.
For the group velocity case,
$\alpha=1$ also introduces a frequency independent phase shift, and therefore  there is an exact degeneracy lensing.
 From Eq.\ (\ref{eq:DeltaPhi_vg}) we can see that this degeneracy will happen for a GW event at redshift $z$, whenever the modified gravity parameter is given by
\begin{equation} 
    \mbA = 2\pi \frac{n_j+\mathfrak{i}}{D_1(z)}\, \qquad (\alpha=1),
\end{equation}
where $\mathfrak{i}$ is an integer and recall $n_j=1/2$ for the type II image considered here. 
This means that for each redshift, there will always exist many choices of ${\mbA}$ that mimics strong lensing with the smallest such value of $\mbA$ given by $\mathfrak{i}=0$. For this value of $\mbA$ and $\alpha=1$, we then have $\Delta\chi^2=0$. 
Therefore if lensing is not considered in the analysis, then for any signal with a nominal ``$>3\sigma$" unlensed GR template mismatch of $\Delta\chi^2 >9$, there would
be a corresponding ``$>3\sigma$" detection of a non zero $\mbA$ and an erroneous falsification of GR.  

Furthermore even for $\alpha\ne 1$, where the frequency dependent MDR phase and arrival time makes the degeneracy with lensing only partial, there is still a range in SNR where the lensed signal would be mistaken as an MDR.    For example for both extremes $\alpha=0,3$ the fits are comparable and better than with the unlensed GR template: $\Delta\chi^2/\rho^2 \approx 0.07$ vs $0.15$.
This means that with the nominal ``$3\sigma$" criteria, for signals with $60 \lesssim \rho^2 \lesssim 130$,
these $\alpha$ cases provide a  good fit to the signal while unlensed GR provides a bad fit.    Hence lensing would be mistaken for these MDRs, with a $3\sigma$ detection of a non zero amplitude $\mbA$ for $\rho^2\gtrsim 110$.
Moreover, we can see how the MDR fits worse the lensing signal as we move away from $\alpha=1$ at both sides, due to the faster frequency dependence of the phase shift induced by the MDRs.

Although both $\alpha=0,3$ fit the total signal nearly as well, they do so for different reasons as shown 
in Fig.\ \ref{fig:chi2-f_lensing} (bottom panel).
For $\alpha=0$, where the MDR changes peak at low frequency, the fit  is likewise optimized to lower half of frequency space in terms of contributions to the total $\rho^2$, which includes  the inspiral regime of the 22 mode, whereas for $\alpha=3$, where the MDR changes peak at high frequency, the fit tries to compromise to accommodate the range above $f_{22,c}$ from the 33 and 44 modes.
Note that this fit is subject to the positioning of $f_{j,c}$ with respect to the minimum of the power spectral density $S_n(f)$, and hence depends on the source masses and redshift.

For a given source we can use $T_{\rm shift}$ to understand these trends with frequency.   
Since each frequency suffers a time delay according to 
Eq.~(\ref{eq:Tjc}), the best fit value of $T_{\rm shift}$ reflects the frequency $f_{\rm eff}$ at which the template attempts to fit the signal the best
\begin{equation}
\frac{ T_{\rm shift} }{T_{22,c}}=
\left( \frac{ f_{\rm eff} }{f_{22,c} }\right)^{\alpha-2}.
\end{equation}
We can verify this interpretation by constructing the arrival phase at this frequency using Eq.~(\ref{eq:DeltaPhi_vg}) as
\begin{equation}
\Delta \Phi(f_{\rm eff}) = \left( \frac{ T_{\rm shift} }{T_{22,c}}\right)^{\frac{\alpha-1}{\alpha-2}}\Delta \Phi(f_{22,c})
\end{equation}
and comparing the phase shift to the lensing phase shift
\begin{equation}
\Delta \Phi(f_{\rm eff}) + 2\Delta\varphi_c \approx \pi/2\,.\label{feff_fit}
\end{equation}
Tab.~\ref{tab:chi2_baseline} shows that this equivalence roughly holds for all $\alpha$ whereas $\Delta\Phi(f_{22,c})+2\Delta \varphi_c$ does not. 
We can also see this in Fig.\ \ref{fig:chi2-A} by looking at the location of the minima.

%--FIG: chi2 vs alpha
\begin{figure}[t!]
\centering
\includegraphics[width=0.7\textwidth]{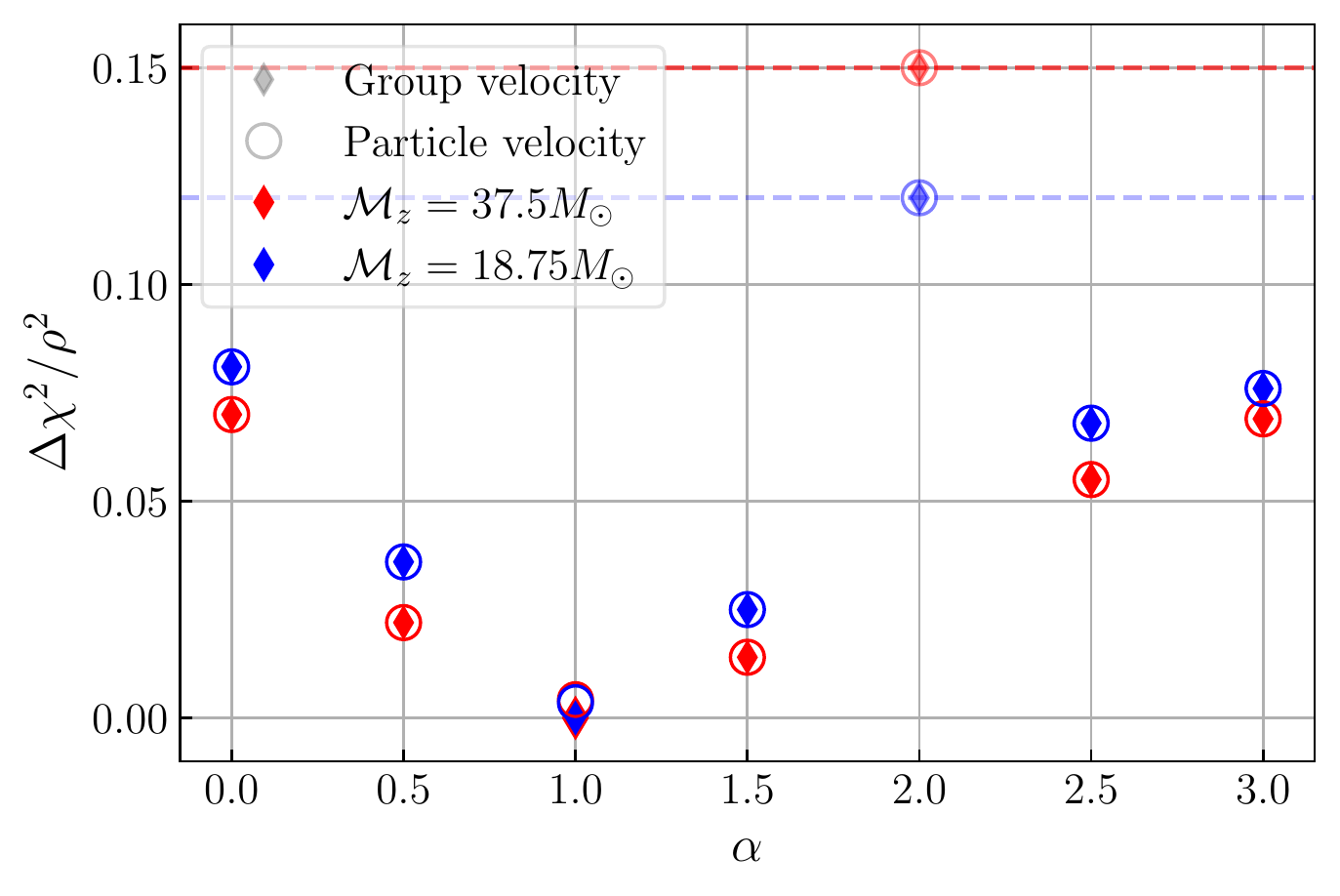}
\caption{\label{fig:chi2-alpha}
Goodness of fit of a lensed type II image with a modified gravity template. We plot the $\Delta\chi^2/\rho^2$ as a function of the power $\alpha$ of the MDR. 
Diamonds correspond to MDRs using the group velocity while circles to MDRs using the particle velocity.
For the modified gravity template we minimize the $\Delta\chi^2$ varying the MDR amplitude $\mbA$, and the coalescence phase and time. We quote the results for different chirp masses and fixed mass ratio $q=0.1$ and inclination $\iota=\pi/3$. 
The horizontal, dashed lines correspond to the fit with an unlensed GR template, whose fit is equivalent to $\alpha=2$. 
For the particle velocity with $\alpha=1$ we use the LVK convention for $f_0$.
}
\end{figure}
%----.

Next we can also test the fit using the standard particle velocity approach. Since the particle velocity modifications for $\alpha\neq1$ only differ from the propagation approach by a constant amplitude rescaling, we do not repeat the minimization in these cases as it would only change the best fit $\mbA$ by $1/(1-\alpha)$ with the replacement $\Delta\Phi \rightarrow \Delta\tilde\Phi$ (see Tab.~\ref{table:notation} for phase definitions). 
Only for $\alpha=1$ there is a difference in the $\Delta\chi^2$ between the group and particle velocity.
For $\alpha=1$, we show the best-fit results in the last row of each fiducial binary Tab.~\ref{tab:chi2_baseline}. 
We also plot $\Delta\chi^2/\rho^2$ as a function of $\Delta\Phi(f_{22,c})$ in the right panel of Fig.\ \ref{fig:chi2-A}. There, we can see the difference in the global minimum between both approaches but, more importantly, the fact that for the group velocity the degeneracy is exact and therefore it shows a periodic behavior with all minima at $\Delta\chi^2=0$, while the particle velocity model has a quasi-periodic behavior due to the increasing amplitude of the logarithmic  frequency evolution of the phase distortion, and thus has local minima beyond $|\Delta\Phi|\sim\pi/2$ that give a worse fit, but still better than GR.
For other values of $\alpha$ in the left panel of Fig.\ \ref{fig:chi2-A}, there is no quasi-periodic behavior in $\Delta\chi^2/\rho^2$ and the fits remain worse than GR beyond the limits shown.

As discussed before, for the particle velocity model we have chosen the arbitrary normalization frequency $f_0$ to match that of the LVK convention. 
As expected, we see that the logarithmic frequency-dependence phase shift of $\alpha=1$ evolves slowly enough with frequency to fit a type II strongly-lensed signal very well, but unlike the group velocity case $\alpha=1$ it does not do it perfectly. For example, for $\mathcal{M}_z=37.5 M_\odot$, $\Delta\chi^2/\rho^2=0.0042$ with $f_0 \approx  66 f_{22,c}$ 
and it would take a signal with $\rho\gtrsim 50$ to be considered a bad fit at the ``$3\sigma$" level.
Finally, it is important to note that the mismatch in $\Delta\chi^2$ depends on the arbitrary value of $f_0$ and it is only in the limit that $f_0$ is much greater than or much less than the observed frequencies that the perfect fit of the group velocity is recovered $\Delta\chi^2\to0$. 
For example, the worst case scenario $\mathcal{M}_z=37.5$ is to fix $f_0 \approx 0.8 f_{22,c}$, in which case $\Delta\chi^2/\rho^2=0.15$ and there is  no improvement over the unlensed GR template.
Although the LVK convention for $f_0$ is unphysical in that it depends on the chirp mass, it has the advantage that $f_0/f_{22,c}\approx 66 \gg 1$ independently of mass.

Our main results for the goodness of fit of a type II image to an MDR with different $\alpha$ are illustrated in Fig. \ref{fig:chi2-alpha} for a fixed
$q=0.1$ and $\iota=\pi/3$, and considering two redshifted chirp masses $\mathcal{M}_z = 37.5$ and $18.75 M_\odot$. 
All the diamonds correspond to models with an MDR using the group velocity approach, whereas the circles indicate MDR with particle velocity
and the fiducial LVK $f_0$ choice.
The dashed lines indicate the fit with an unlensed GR template. If the MDR cases lie significantly below this line there is a range of SNR values where the lensed signal may be misidentified as an MDR.  
In this plot we can see that the case of $\alpha=1$ yields the best results, 
since the phase shift is independent of frequency in the group velocity case and very weakly dependent on frequency in the particle velocity case since $f_0\gg f_{22c}$.
For $\alpha=2$, the MDR changes only the arrival time of the signal and so once $T_{\rm shift}$ is optimized, the templates fit exactly the same as in GR.
 For $\alpha\ne 1,2$ and these types of sources, we can  see that for higher chirp masses, the relative improvement of the MDR fit over unlensed GR is larger, due to the fact that higher mass events have a shorter detectable frequency range, and thus the frequency dependent MDR phase correction can mimic better the frequency-independent lensing phase shift. This  trend with mass depends on 
 the detectability of the higher modes across the high SNR frequency range, and thus depends on both the parameters of the source and the detector, as we have discussed in Fig.\ \ref{fig:chi2-q}.

\section{Conclusions}\label{sec:conclusions}

Gravitational wave (GW) catalogs are rapidly increasing in number, with more events having relevant weight on their higher modes. 
We have studied how a generic modified GW propagation could distort such signals. 
We have shown how the propagation (WKB) approach \cite{Ezquiaga:2021ler} is independent of the emitted signal and thus could be directly applied to \emph{any} type of signal to predict analytically the phase distortions in frequency-domain. 
Then we compared this WKB approach  with applying a stationary phase approximation (SPA) on the temporal waveform, which has been a common route in the literature, e.g.\ \cite{Will:1997bb,Mirshekari:2011yq,LIGOScientific:2019fpa}. 
We have demonstrated that the WKB approach is equivalent to the SPA, once the reference phases and arrival times for each mode of the latter are consistently treated, as summarized in Tab.~\ref{table:notation},
and that this naturally determines the \emph{group velocity} as the propagation speed of the wave. 
This is however different from the result when assuming that GWs propagate at the \emph{particle velocity}, for any MDR of the form $\omega^2- c^2k^2 \propto k^\alpha$, especially for $\alpha=1$ where the two differ in functional form for the frequency dependent phase. In this paper, we compared both the group and particle velocity approaches, and demonstrated  how to translate constraints between both approaches for GWs with multiple frequency modes.
Our WKB formalism will be relevant for future analyses of a modified GW propagation using templates with higher modes.

In addition, we analyzed how astrophysical phenomena within GR could be confused with a modified GW propagation in the presence of higher modes.
In particular, we focused on the case of strong gravitational lensing, since each lensed GW image acquires a frequency independent phase shift.  Whereas for each mode of order $m$, this constant phase shift is degenerate with the coalescence phase $\varphi_c$, the presence of higher modes beyond $|m|=2$ breaks this degeneracy for images of type II.
For example, for a GW with signal-to-noise ratio (SNR) $\rho>8$ in an A+ LIGO detector \cite{Aasi:2013wya}, and mass ratio $q=m_2/m_1\lesssim0.1$ and for the 50\% of sources with inclination $\iota\gtrsim0.3\pi$, a typical fit using unlensed GR templates would have a $\Delta\chi^2>9$, indicating ``$>3\sigma$'' bad fit/evidence for modified gravity, assuming that the source-detector Euler angles $(\theta,\phi,\psi)$ can be disentangled with multiply oriented detectors. 
If the Euler angles cannot be measured independently, the significance of this apparent GR violation will be degraded due to the partial degeneracies of the GW phase with these parameters (see \cite{Ezquiaga:2020gdt}, Fig.~8 for the degradation due to $\psi$ as a function of $q$ for a single detector).

In addition, we found that the lensing phase shift is  exactly degenerate with an MDR with corrections linear in frequency, i.e.\ $\omega^2- c^2k^2 \propto k$.
In this case, strong lensing can always be misidentified as an MDR if the higher modes are significant enough to break the $\varphi_c$ degeneracy.
We also quantify the level of similarity for other types of dispersion relations ($\alpha\ne1$) by computing the goodness-of-fit $\Delta\chi^2$ between a modified gravity template and a strongly lensed signal. 
We find that for the approximate degeneracies ($\alpha\neq1$) there is a SNR range where an MDR template would give a better fit than an unlensed GR template, see Fig. \ref{fig:chi2-alpha}, and therefore the lensed signal could be misidentified as modified gravity. These degeneracies are however broken at high SNRs, $\rho \gtrsim 25$ for asymmetric binaries with $q\lesssim0.1$ and $\iota\gtrsim0.3\pi$. 
Instead, for the particle velocity parametrization of $\alpha=1$ used by the LIGO--Virgo--KAGRA collaboration, one would need SNRs $\rho \gtrsim 50$ for the same type of binary parameters to break the MDR degeneracy with strong lensing, since this MDR model introduces a phase correction with a slow logarithmic running in frequency. 
For such large SNRs, waveform systematics may start to become relevant, possibly leading to other sources of spurious detection of modified gravity. Such effects should be considered in future analyses. 
Moreover, our detectability criteria from the $\Delta\chi^2$ varying only a few template parameters could be improved performing a full parameter estimation.

Finally, we emphasize that these degeneracies between modified gravity and strong lensing are mostly applicable on an event by event basis. 
When considering the whole population, the potential observed deviations from GR should be consistent among each other, something difficult to mimic if the events had indeed been lensed. 
Still, as we have demonstrated here, current pipelines searching for modified GW propagation might misinterpret lensed events with modified gravity. 
We have shown how to identify if this apparent modification of gravity is indeed a lensed signal, although we advocate for the simpler solution of applying type II image searches in the data. 
This could be achieved simultaneously if each modified gravity parameter point automatically maximizes the likelihood over the discrete $0,\pi/2,\pi$ lensing phase shifts for each set of parameters.

%Acknowledgments 
\section*{Acknowledgments}

We are grateful to Keisuke Inomata, Max Isi, Austin Joyce, Hayden Lee for useful discussions and Alex Nitz for correspondence about \texttt{pyCBC} \cite{alex_nitz_2019_3546372}, which we use to generate the GR waveforms. 
JME is supported by NASA through the NASA Hubble Fellowship grant HST-HF2-51435.001-A awarded by the Space Telescope Science Institute, which is operated by the Association of Universities for Research in Astronomy, Inc., for NASA, under contract NAS5-26555. He is also supported by the Kavli Institute for Cosmological Physics through an endowment from the Kavli Foundation and its founder Fred Kavli. WH and MXL were supported by the U.S.~Dept.~of Energy contract DE-FG02-13ER41958 and the Simons Foundation. ML was supported by the Innovative Theory Cosmology Fellowship at the University of Columbia. FX gratefully acknowledges the Elaine K. Bernstein Award.

%-----------------
%-----------------
%Bib
\bibliographystyle{JHEP}
\bibliography{gw_phase_degeneracies}

\providecommand{\href}[2]{#2}\begingroup\raggedright\begin{thebibliography}{10}

\bibitem{Aasi2015}
J.~A. et~al., \emph{Advanced {LIGO}},
  \href{http://dx.doi.org/10.1088/0264-9381/32/7/074001}{\emph{Classical and
  Quantum Gravity} {\bfseries 32} (mar, 2015) 074001}.

\bibitem{Acernese_2014}
F.~A. et~al, \emph{Advanced virgo: a second-generation interferometric
  gravitational wave detector},
  \href{http://dx.doi.org/10.1088/0264-9381/32/2/024001}{\emph{Classical and
  Quantum Gravity} {\bfseries 32} (dec, 2014) 024001}.

\bibitem{KAGRA:2020tym}
{\scshape KAGRA} collaboration, T.~Akutsu et~al., \emph{{Overview of KAGRA:
  Detector design and construction history}},
  \href{http://dx.doi.org/10.1093/ptep/ptaa125}{\emph{PTEP} {\bfseries 2021}
  (2021) 05A101}, [\href{https://arxiv.org/abs/2005.05574}{{\ttfamily
  2005.05574}}].

\bibitem{LIGOScientific:2021djp}
{\scshape LIGO Scientific, VIRGO, KAGRA} collaboration, R.~Abbott et~al.,
  \emph{{GWTC-3: Compact Binary Coalescences Observed by LIGO and Virgo During
  the Second Part of the Third Observing Run}},
  \href{https://arxiv.org/abs/2111.03606}{{\ttfamily 2111.03606}}.

\bibitem{LIGOScientific:2020stg}
{\scshape LIGO Scientific, Virgo} collaboration, R.~Abbott et~al.,
  \emph{{GW190412: Observation of a Binary-Black-Hole Coalescence with
  Asymmetric Masses}},  \href{https://arxiv.org/abs/2004.08342}{{\ttfamily
  2004.08342}}.

\bibitem{Abbott:2020khf}
{\scshape LIGO Scientific, Virgo} collaboration, R.~Abbott et~al.,
  \emph{{GW190814: Gravitational Waves from the Coalescence of a 23 Solar Mass
  Black Hole with a 2.6 Solar Mass Compact Object}},
  \href{http://dx.doi.org/10.3847/2041-8213/ab960f}{\emph{Astrophys. J.}
  {\bfseries 896} (2020) L44},
  [\href{https://arxiv.org/abs/2006.12611}{{\ttfamily 2006.12611}}].

\bibitem{Olsen:2022pin}
S.~Olsen, T.~Venumadhav, J.~Mushkin, J.~Roulet, B.~Zackay and M.~Zaldarriaga,
  \emph{{New binary black hole mergers in the LIGO--Virgo O3a data}},
  \href{https://arxiv.org/abs/2201.02252}{{\ttfamily 2201.02252}}.

\bibitem{Nitz:2021zwj}
A.~H. Nitz, S.~Kumar, Y.-F. Wang, S.~Kastha, S.~Wu, M.~Sch\"afer et~al.,
  \emph{{4-OGC: Catalog of gravitational waves from compact-binary mergers}},
  \href{https://arxiv.org/abs/2112.06878}{{\ttfamily 2112.06878}}.

\bibitem{LISA:2017pwj}
{\scshape LISA} collaboration, P.~Amaro-Seoane et~al., \emph{{Laser
  Interferometer Space Antenna}},
  \href{https://arxiv.org/abs/1702.00786}{{\ttfamily 1702.00786}}.

\bibitem{Maggiore:2019uih}
M.~Maggiore et~al., \emph{{Science Case for the Einstein Telescope}},
  \href{http://dx.doi.org/10.1088/1475-7516/2020/03/050}{\emph{JCAP} {\bfseries
  03} (2020) 050}, [\href{https://arxiv.org/abs/1912.02622}{{\ttfamily
  1912.02622}}].

\bibitem{Evans:2021gyd}
M.~Evans et~al., \emph{{A Horizon Study for Cosmic Explorer: Science,
  Observatories, and Community}},
  \href{https://arxiv.org/abs/2109.09882}{{\ttfamily 2109.09882}}.

\bibitem{Dai:2017huk}
L.~Dai and T.~Venumadhav, \emph{{On the waveforms of gravitationally lensed
  gravitational waves}},  \href{https://arxiv.org/abs/1702.04724}{{\ttfamily
  1702.04724}}.

\bibitem{Ezquiaga:2020gdt}
J.~M. Ezquiaga, D.~E. Holz, W.~Hu, M.~Lagos and R.~M. Wald, \emph{{Phase
  effects from strong gravitational lensing of gravitational waves}},
  \href{http://dx.doi.org/10.1103/PhysRevD.103.064047}{\emph{Phys. Rev. D}
  {\bfseries 103} (2021) 064047},
  [\href{https://arxiv.org/abs/2008.12814}{{\ttfamily 2008.12814}}].

\bibitem{Yunes:2009ke}
N.~Yunes and F.~Pretorius, \emph{{Fundamental Theoretical Bias in Gravitational
  Wave Astrophysics and the Parameterized Post-Einsteinian Framework}},
  \href{http://dx.doi.org/10.1103/PhysRevD.80.122003}{\emph{Phys. Rev.}
  {\bfseries D80} (2009) 122003},
  [\href{https://arxiv.org/abs/0909.3328}{{\ttfamily 0909.3328}}].

\bibitem{Tahura:2018zuq}
S.~Tahura and K.~Yagi, \emph{{Parameterized Post-Einsteinian Gravitational
  Waveforms in Various Modified Theories of Gravity}},
  \href{http://dx.doi.org/10.1103/PhysRevD.98.084042}{\emph{Phys. Rev. D}
  {\bfseries 98} (2018) 084042},
  [\href{https://arxiv.org/abs/1809.00259}{{\ttfamily 1809.00259}}].

\bibitem{Jimenez:2019lrk}
J.~B. Jim\'enez, J.~M. Ezquiaga and L.~Heisenberg, \emph{{Probing cosmological
  fields with gravitational wave oscillations}},
  \href{http://dx.doi.org/10.1088/1475-7516/2020/04/027}{\emph{JCAP} {\bfseries
  04} (2020) 027}, [\href{https://arxiv.org/abs/1912.06104}{{\ttfamily
  1912.06104}}].

\bibitem{Will:1997bb}
C.~M. Will, \emph{{Bounding the mass of the graviton using gravitational wave
  observations of inspiralling compact binaries}},
  \href{http://dx.doi.org/10.1103/PhysRevD.57.2061}{\emph{Phys. Rev.}
  {\bfseries D57} (1998) 2061--2068},
  [\href{https://arxiv.org/abs/gr-qc/9709011}{{\ttfamily gr-qc/9709011}}].

\bibitem{Mirshekari:2011yq}
S.~Mirshekari, N.~Yunes and C.~M. Will, \emph{{Constraining Generic Lorentz
  Violation and the Speed of the Graviton with Gravitational Waves}},
  \href{http://dx.doi.org/10.1103/PhysRevD.85.024041}{\emph{Phys. Rev. D}
  {\bfseries 85} (2012) 024041},
  [\href{https://arxiv.org/abs/1110.2720}{{\ttfamily 1110.2720}}].

\bibitem{LIGOScientific:2019fpa}
{\scshape LIGO Scientific, Virgo} collaboration, B.~P. Abbott et~al.,
  \emph{{Tests of General Relativity with the Binary Black Hole Signals from
  the LIGO-Virgo Catalog GWTC-1}},
  \href{http://dx.doi.org/10.1103/PhysRevD.100.104036}{\emph{Phys. Rev.}
  {\bfseries D100} (2019) 104036},
  [\href{https://arxiv.org/abs/1903.04467}{{\ttfamily 1903.04467}}].

\bibitem{Ezquiaga:2021ler}
J.~M. Ezquiaga, W.~Hu, M.~Lagos and M.-X. Lin, \emph{{Gravitational wave
  propagation beyond general relativity: waveform distortions and echoes}},
  \href{http://dx.doi.org/10.1088/1475-7516/2021/11/048}{\emph{JCAP} {\bfseries
  11} (2021) 048}, [\href{https://arxiv.org/abs/2108.10872}{{\ttfamily
  2108.10872}}].

\bibitem{Abbott:2020jks}
{\scshape LIGO Scientific, Virgo} collaboration, R.~Abbott et~al., \emph{{Tests
  of general relativity with binary black holes from the second LIGO-Virgo
  gravitational-wave transient catalog}},
  \href{http://dx.doi.org/10.1103/PhysRevD.103.122002}{\emph{Phys. Rev. D}
  {\bfseries 103} (2021) 122002},
  [\href{https://arxiv.org/abs/2010.14529}{{\ttfamily 2010.14529}}].

\bibitem{LIGOScientific:2021sio}
{\scshape LIGO Scientific, VIRGO, KAGRA} collaboration, R.~Abbott et~al.,
  \emph{{Tests of General Relativity with GWTC-3}},
  \href{https://arxiv.org/abs/2112.06861}{{\ttfamily 2112.06861}}.

\bibitem{Aasi:2013wya}
{\scshape KAGRA, LIGO Scientific, VIRGO} collaboration, B.~P. Abbott et~al.,
  \emph{{Prospects for Observing and Localizing Gravitational-Wave Transients
  with Advanced LIGO, Advanced Virgo and KAGRA}},
  \href{http://dx.doi.org/10.1007/s41114-018-0012-9,
  10.1007/lrr-2016-1}{\emph{Living Rev. Rel.} {\bfseries 21} (2018) 3},
  [\href{https://arxiv.org/abs/1304.0670}{{\ttfamily 1304.0670}}].

\bibitem{2020JCAP...03..050M}
M.~{Maggiore}, C.~{Van Den Broeck}, N.~{Bartolo}, E.~{Belgacem}, D.~{Bertacca},
  M.~A. {Bizouard} et~al., \emph{{Science case for the Einstein telescope}},
  \href{http://dx.doi.org/10.1088/1475-7516/2020/03/050}{\emph{Journal of
  Cosmology and Astroparticle Physics} {\bfseries 2020} (Mar., 2020) 050},
  [\href{https://arxiv.org/abs/1912.02622}{{\ttfamily 1912.02622}}].

\bibitem{2019BAAS...51g..35R}
D.~{Reitze}, R.~X. {Adhikari}, S.~{Ballmer}, B.~{Barish}, L.~{Barsotti},
  G.~{Billingsley} et~al., \emph{{Cosmic Explorer: The U.S. Contribution to
  Gravitational-Wave Astronomy beyond LIGO}},  in \emph{Bulletin of the
  American Astronomical Society}, vol.~51, p.~35, Sept., 2019,
  \href{https://arxiv.org/abs/1907.04833}{{\ttfamily 1907.04833}}.

\bibitem{2018MNRAS.476.2220L}
S.-S. {Li}, S.~{Mao}, Y.~{Zhao} and Y.~{Lu}, \emph{{Gravitational lensing of
  gravitational waves: a statistical perspective}},
  \href{http://dx.doi.org/10.1093/mnras/sty411}{\emph{Monthly Notices of the
  Royal Astronomical Society} {\bfseries 476} (May, 2018) 2220--2229},
  [\href{https://arxiv.org/abs/1802.05089}{{\ttfamily 1802.05089}}].

\bibitem{2019arXiv190403187T}
{The LIGO Scientific collaboration}, \emph{{Gravitational wave astronomy with
  LIGO and similar detectors in the next decade}}, {\emph{arXiv e-prints}
  (Apr., 2019) arXiv:1904.03187},
  [\href{https://arxiv.org/abs/1904.03187}{{\ttfamily 1904.03187}}].

\bibitem{2021arXiv210514390X}
F.~{Xu}, J.~M. {Ezquiaga} and D.~E. {Holz}, \emph{{Please repeat: Strong
  lensing of gravitational waves as a probe of compact binary and galaxy
  populations}}, {\emph{arXiv e-prints} (May, 2021) arXiv:2105.14390},
  [\href{https://arxiv.org/abs/2105.14390}{{\ttfamily 2105.14390}}].

\bibitem{Caliskan:2022wbh}
M.~\c{C}al\i{}\c{s}kan, J.~M. Ezquiaga, O.~A. Hannuksela and D.~E. Holz,
  \emph{{Lensing or luck? False alarm probabilities for gravitational lensing
  of gravitational waves}},  \href{https://arxiv.org/abs/2201.04619}{{\ttfamily
  2201.04619}}.

\bibitem{Jim_nez_2020}
J.~B. Jim{\'{e}}nez, J.~M. Ezquiaga and L.~Heisenberg, \emph{Probing
  cosmological fields with gravitational wave oscillations},
  \href{http://dx.doi.org/10.1088/1475-7516/2020/04/027}{\emph{Journal of
  Cosmology and Astroparticle Physics} {\bfseries 2020} (apr, 2020) 027--027}.

\bibitem{Yunes:2009yz}
N.~Yunes, K.~Arun, E.~Berti and C.~M. Will, \emph{{Post-Circular Expansion of
  Eccentric Binary Inspirals: Fourier-Domain Waveforms in the Stationary Phase
  Approximation}},
  \href{http://dx.doi.org/10.1103/PhysRevD.80.084001}{\emph{Phys. Rev. D}
  {\bfseries 80} (2009) 084001},
  [\href{https://arxiv.org/abs/0906.0313}{{\ttfamily 0906.0313}}].

\bibitem{Cutler:1994ys}
C.~Cutler and E.~E. Flanagan, \emph{{Gravitational waves from merging compact
  binaries: How accurately can one extract the binary's parameters from the
  inspiral wave form?}},
  \href{http://dx.doi.org/10.1103/PhysRevD.49.2658}{\emph{Phys. Rev. D}
  {\bfseries 49} (1994) 2658--2697},
  [\href{https://arxiv.org/abs/gr-qc/9402014}{{\ttfamily gr-qc/9402014}}].

\bibitem{Droz:1999qx}
S.~Droz, D.~J. Knapp, E.~Poisson and B.~J. Owen, \emph{{Gravitational waves
  from inspiraling compact binaries: Validity of the stationary phase
  approximation to the Fourier transform}},
  \href{http://dx.doi.org/10.1103/PhysRevD.59.124016}{\emph{Phys. Rev. D}
  {\bfseries 59} (1999) 124016},
  [\href{https://arxiv.org/abs/gr-qc/9901076}{{\ttfamily gr-qc/9901076}}].

\bibitem{Maggiore:1900zz}
M.~Maggiore, \emph{{Gravitational Waves. Vol. 1: Theory and Experiments}}.
\newblock Oxford Master Series in Physics. Oxford University Press, 2007.

\bibitem{LIGOScientific:2017vwq}
{\scshape LIGO Scientific, Virgo} collaboration, B.~P. Abbott et~al.,
  \emph{{GW170817: Observation of Gravitational Waves from a Binary Neutron
  Star Inspiral}},
  \href{http://dx.doi.org/10.1103/PhysRevLett.119.161101}{\emph{Phys. Rev.
  Lett.} {\bfseries 119} (2017) 161101},
  [\href{https://arxiv.org/abs/1710.05832}{{\ttfamily 1710.05832}}].

\bibitem{Mastrogiovanni:2020gua}
S.~Mastrogiovanni, D.~Steer and M.~Barsuglia, \emph{{Probing modified gravity
  theories and cosmology using gravitational-waves and associated
  electromagnetic counterparts}},
  \href{http://dx.doi.org/10.1103/PhysRevD.102.044009}{\emph{Phys. Rev. D}
  {\bfseries 102} (2020) 044009},
  [\href{https://arxiv.org/abs/2004.01632}{{\ttfamily 2004.01632}}].

\bibitem{Baker:2022rhh}
T.~Baker et~al., \emph{{Measuring the propagation speed of gravitational waves
  with LISA}},  \href{https://arxiv.org/abs/2203.00566}{{\ttfamily
  2203.00566}}.

\bibitem{PhysRevLett.120.161102}
L.~London, S.~Khan, E.~Fauchon-Jones, C.~Garc\'{\i}a, M.~Hannam, S.~Husa
  et~al., \emph{First higher-multipole model of gravitational waves from
  spinning and coalescing black-hole binaries},
  \href{http://dx.doi.org/10.1103/PhysRevLett.120.161102}{\emph{Phys. Rev.
  Lett.} {\bfseries 120} (Apr, 2018) 161102}.

\bibitem{Wang:2021kzt}
Y.~Wang, R.~K.~L. Lo, A.~K.~Y. Li and Y.~Chen, \emph{{Identifying Type-II
  Strongly-Lensed Gravitational-Wave Images in Third-Generation
  Gravitational-Wave Detectors}},
  \href{https://arxiv.org/abs/2101.08264}{{\ttfamily 2101.08264}}.

\bibitem{Janquart:2021nus}
J.~Janquart, E.~Seo, O.~A. Hannuksela, T.~G.~F. Li and C.~V.~D. Broeck,
  \emph{{On the Identification of Individual Gravitational-wave Image Types of
  a Lensed System Using Higher-order Modes}},
  \href{http://dx.doi.org/10.3847/2041-8213/ac3bcf}{\emph{Astrophys. J. Lett.}
  {\bfseries 923} (2021) L1},
  [\href{https://arxiv.org/abs/2110.06873}{{\ttfamily 2110.06873}}].

\bibitem{Vijaykumar:2022dlp}
A.~Vijaykumar, A.~K. Mehta and A.~Ganguly, \emph{{Detection and parameter
  estimation challenges of Type-II lensed binary black hole signals}},
  \href{https://arxiv.org/abs/2202.06334}{{\ttfamily 2202.06334}}.

\bibitem{Dai:2020tpj}
L.~Dai, B.~Zackay, T.~Venumadhav, J.~Roulet and M.~Zaldarriaga, \emph{{Search
  for Lensed Gravitational Waves Including Morse Phase Information: An
  Intriguing Candidate in O2}},
  \href{https://arxiv.org/abs/2007.12709}{{\ttfamily 2007.12709}}.

\bibitem{Liu:2020par}
X.~Liu, I.~M. Hernandez and J.~Creighton, \emph{{Identifying strong
  gravitational-wave lensing during the second observing run of Advanced LIGO
  and Advanced Virgo}},
  \href{http://dx.doi.org/10.3847/1538-4357/abd7eb}{\emph{Astrophys. J.}
  {\bfseries 908} (2021) 97},
  [\href{https://arxiv.org/abs/2009.06539}{{\ttfamily 2009.06539}}].

\bibitem{LIGOScientific:2021izm}
{\scshape LIGO Scientific, VIRGO} collaboration, R.~Abbott et~al.,
  \emph{{Search for Lensing Signatures in the Gravitational-Wave Observations
  from the First Half of LIGO\textendash{}Virgo\textquoteright{}s Third
  Observing Run}},
  \href{http://dx.doi.org/10.3847/1538-4357/ac23db}{\emph{Astrophys. J.}
  {\bfseries 923} (2021) 14},
  [\href{https://arxiv.org/abs/2105.06384}{{\ttfamily 2105.06384}}].

\bibitem{Lo:2021nae}
R.~K.~L. Lo and I.~Maga\~na Hernandez, \emph{{A Bayesian statistical framework
  for identifying strongly-lensed gravitational-wave signals}},
  \href{https://arxiv.org/abs/2104.09339}{{\ttfamily 2104.09339}}.

\bibitem{Janquart:2021qov}
J.~Janquart, O.~A. Hannuksela, H.~K. and C.~Van Den~Broeck, \emph{{A fast and
  precise methodology to search for and analyse strongly lensed
  gravitational-wave events}},
  \href{http://dx.doi.org/10.1093/mnras/stab1991}{\emph{Mon. Not. Roy. Astron.
  Soc.} {\bfseries 506} (2021) 5430--5438},
  [\href{https://arxiv.org/abs/2105.04536}{{\ttfamily 2105.04536}}].

\bibitem{Lindblom:2008cm}
L.~Lindblom, B.~J. Owen and D.~A. Brown, \emph{{Model Waveform Accuracy
  Standards for Gravitational Wave Data Analysis}},
  \href{http://dx.doi.org/10.1103/PhysRevD.78.124020}{\emph{Phys. Rev. D}
  {\bfseries 78} (2008) 124020},
  [\href{https://arxiv.org/abs/0809.3844}{{\ttfamily 0809.3844}}].

\bibitem{alex_nitz_2019_3546372}
A.~Nitz, I.~Harry, D.~Brown, C.~M. Biwer, J.~Willis, T.~D. Canton et~al.,
  \emph{gwastro/pycbc: Pycbc release v1.14.4},  Nov., 2019.
\newblock 10.5281/zenodo.3546372.

\bibitem{LIGOScientific:2016ebw}
{\scshape LIGO Scientific, Virgo} collaboration, B.~P. Abbott et~al.,
  \emph{{Effects of waveform model systematics on the interpretation of
  GW150914}}, \href{http://dx.doi.org/10.1088/1361-6382/aa6854}{\emph{Class.
  Quant. Grav.} {\bfseries 34} (2017) 104002},
  [\href{https://arxiv.org/abs/1611.07531}{{\ttfamily 1611.07531}}].

\end{thebibliography}\endgroup
%------
\end{document}